\documentclass[aps,pra,reprint,nofootinbib,superscriptaddress]{revtex4-1}
\usepackage[mathscr]{eucal}
\usepackage[english]{babel}
\usepackage{amsmath,amsxtra,amssymb, amstext}
\usepackage{mathtools}
\usepackage{extarrows}
\usepackage{dsfont}
\usepackage{footnote}	 					 
\usepackage{physics}
\usepackage[colorlinks=true, linkcolor =blue, citecolor=green!70!black]{hyperref}
\usepackage{natbib}
\usepackage{xcolor}
\usepackage{scalerel}
\usepackage{cleveref}
\usepackage{tikz}
\usetikzlibrary{positioning}
\DeclareRobustCommand{\us}{\bgroup\markunderlinestart}
\newcommand{\markunderlinestart}[1]{\underline{#1}\egroup}
\newcommand{\mo}[1]{\mathcal{O}_{#1}}
\newcommand{\mot}[1]{\mathcal{O}_{#1}^\mathrm{th}}
\newcommand{\mog}[1]{\mathcal{O}_{#1}^{\mathrm{g} }}
\newcommand{\sm}[1]{L_{#1}^{\phantom{i}#1'}}
\newcommand{\im}{\mathrm{i}}
\newcommand{\E}{\mathrm{e}}
\newcommand{\veca}{\hat{\vec{a}}}

\newcommand{\cov}[1]{\mathrm{cov}_{#1}}

\usepackage{soul}

\begin{document}

\title{Exploiting higher-order correlation functions for photon-statistics-based characterization and reconstruction of arbitrary Gaussian states}

\author{Philip Heinzel}
\email{philip.heinzel@uni-jena.de}
\affiliation{Institute of Condensed Matter Theory and Optics, Friedrich-Schiller-University Jena, Max-Wien-Platz 1, 07743 Jena, Germany}

\author{René Sondenheimer}
\email{rene.sondenheimer@uni-jena.de}
\affiliation{Institute of Condensed Matter Theory and Optics, Friedrich-Schiller-University Jena, Max-Wien-Platz 1, 07743 Jena, Germany}
\affiliation{Fraunhofer Institute for Applied Optics and Precision Engineering IOF, Albert-Einstein-Str. 7, 07745 Jena, Germany}

\begin{abstract}
Gaussian states are an essential building block for various applications in quantum optics and quantum information science, yet the precise relation between their second- and third-order correlation functions remains not fully explored. We discuss connections between these correlation functions by constructing an explicit decomposition formula for arbitrary sixth-order moments of ladder operators for general Gaussian states and demonstrate how the derived relations enable state classification from correlation data alone. Whereas violating these relations certifies non-Gaussianity, satisfying them provides evidence for a Gaussian-state description and allows a direct distinction among non-displaced, non-squeezed, and displaced-squeezed sectors of the Gaussian state space. Further, we show that it is not possible to uniquely extract state parameters solely from correlation-function measurements without prior assumptions about the Gaussian state. Resolving this ambiguity requires additional loss-sensitive information, e.g., measuring the mean intensity or the vacuum overlap of each mode. In particular, we show under which circumstances these measurements can be used to reconstruct a generic Gaussian state.
\end{abstract}

\maketitle

\section{Introduction}
Quantum technologies have rapidly transitioned from theoretical concepts to laboratory and even field demonstrations over the past two decades. Central to many of these advances are Gaussian states owing to their relative ease of generation and manipulation \cite{braunstein_quantum_2005,adesso_entanglement_2007,weedbrook_gaussian_2012}. 
Moreover, Gaussian states offer both analytical tractability and valuable quantum properties, such as multimode entanglement, noise below the standard quantum limit, and sub-Poissonian photon statistics. 
Thus, such states underpin protocols ranging from quantum key distribution and teleportation \cite{grosshans_continuous_2002,usenko_entanglementbased_2014,ruppert_longdistance_2014,filip_measurement_2005,zhao_enhancing_2023,romanrodriguez_multimode_2024,oruganti_continuousvariable_2025} as well as quantum sensing \cite{tan_quantum_2008,chua_quantum_2014,vahlbruch_detection_2016,nichols_multiparameter_2018,mcculler_frequencydependent_2020}, to Gaussian boson sampling and continuous-variable quantum computing \cite{jr_continuous_2004,hamilton_gaussian_2017,madsen_quantum_2022,yu_universal_2023}. They also serve as a basic resource for generating further complex quantum states \cite{grimsmo_quantum_2021,sturges_quantum_2021,pan_protecting_2023,li_method_2024,krstic_hybrid_2024}. Therefore, a comprehensive understanding of their statistical properties and correlations is essential for both fundamental studies and practical applications.
Given their significance, extensive investigations have been dedicated to their description and properties over the years \cite{yeoman_twomode_1993,yuen_twophoton_1976,schumaker_new_1985,schumaker_quantum_1986,zhang_coherent_1990,ma_multimode_1990,simon_quantumnoise_1994,simon_pereshorodecki_2000,duan_inseparability_2000,eisert_introduction_2003,adesso_extremal_2004,eisert_distilling_2002,ferraro_gaussian_2005,vallone_means_2019,brask_gaussian_2022,cardin_photonnumber_2023,fitzke_simulating_2023,mizuno_experimental_2005,lemonde_antibunching_2014}.

To characterize Gaussian states, a wide range of methods has been developed, starting from pioneering work on single-mode systems by Yuen \cite{yuen_twophoton_1976}, extensions to two-mode systems by Caves and Schumaker \cite{schumaker_new_1985,schumaker_quantum_1986}, and further advancements to general multimode systems \cite{ma_multimode_1990,vallone_means_2019}. 
Reconstructing a Gaussian state reduces to determining its first and second moments which uniquely fix its mean vector and covariance matrix. This can be achieved by standard continuous-variable tomography being based on phase-sensitive measurements, e.g., balanced homodyne or heterodyne detection. 
Recent advances in efficient Gaussian-state tomography based on homodyne detection plus adaptive Gaussian operations further showed that sample complexity is (nearly) energy independent \cite{bittel_energyindependent_2025}.
While informationally complete, this route is experimentally sophisticated. It requires a phase-stable local oscillator, careful phase calibration across modes, and large data volumes to suppress statistical noise \cite{lvovsky_continuousvariable_2009}. Losses and mode mismatch may also bias the reconstruction. 
Moreover, determining first- and second-order moments alone does not certify Gaussianity. However, higher-order correlations are uniquely fixed by lower-order moments for Gaussian states, providing nontrivial consistency checks.

These challenges motivate efficient, phase-insensitive diagnostics that exploit the structure of Gaussian states.
A key aspect with that respect are higher-order intensity correlation functions, which encode nonclassical features, e.g., photon antibunching, squeezing, and entanglement. While second-order correlations, $g^{(2)}$, are routinely used to characterize basic properties of the photon-number statistics, third-order correlations, $g^{(3)}$, and higher offer richer information about multiphoton processes. In particular, relations have been derived between $g^{(2)}$ and $g^{(3)}$ coincidence measurements which follow a particular pattern for specific states \cite{christ_probing_2011,wakui_ultrabroadband_2014}. 
Such patterns could be used in the spirit of hypothesis testing to identify certain state characteristics. To date, however, the relationship between $g^{(2)}$ and $g^{(3)}$ has only been explored in special cases, e.g., for thermal light, single-, or uncorrelated multimode squeezed vacuum \cite{wakui_ultrabroadband_2014}. Generalizations of such relations will offer a valuable tool for efficient Gaussian state characterization, e.g., for potential novel sources of quantum light \cite{gonoskov_nonclassical_2022,theidel_evidence_2024,theidel_observation_2024,lange_electroncorrelationinduced_2024,riveradean_squeezed_2024}.

In this work, we close this gap by deriving general analytical connections between $g^{(2)}$ and $g^{(3)}$ for arbitrary multimode Gaussian states. 
Our approach is based on a decomposition formula for sixth-order moments of bosonic ladder operators allowing us to express
third-order photon-counting statistics directly in terms of first- and second-order moments. This unifies and recovers known results for thermal states and single-beam squeezers as limiting cases, while providing a flexible framework for analyzing more complex (multimode) scenarios. We show how the derived relations can be used to distinguish among the classes of non-displaced, non-squeezed, and displaced squeezed Gaussian states.

As $g^{(2)}$ and $g^{(3)}$ are normalized correlation functions, the method is comparatively robust to loss. Further, it does not require a stable phase reference. However, a crucial result of our work is given by the fact that performing such loss-invariant measurement schemes does not enable to uniquely extract certain state parameters on a quantitative level such as squeezing amplitudes and displacements. Estimating state parameters explicitly requires knowledge about a loss-sensitive observable, e.g., the mean photon number that can be acquired by a comparatively simple intensity measurement. Alternatively, the vacuum overlap of the Gaussian state could be used to resolve this ambiguity. Based on these findings, we propose practical schemes for Gaussian state reconstruction by combining second- and third-order correlation functions with loss-sensitive observables. In particular, we show how passive linear optics and intensity measurements or click detectors can be used to reconstruct a generic Gaussian state up to potential reflection symmetries. These methods offer an experimentally accessible route to parameter estimation without full state tomography.

The remainder of the paper is organized as follows. In Sec.~\ref{sec:GaussianStates}, we introduce our conventions and notation by briefly reviewing the parametrization of Gaussian states. 
In Sec.~\ref{sec:CorrelationFunctions}, we construct an explicit decomposition for arbitrary sixth-order ladder operator moments for general Gaussian states. 
This result forms the basis for decomposing the third-order intensity correlation function into lower-order moments allowing us to establish relations between $g^{(2)}$ and $g^{(3)}$. How these relations can be used for verification and a convenient classification of single-mode Gaussian states by loss-invariant measurements is discussed in Sec.~\ref{sec:SingleMode}. Further, we demonstrate that a generic single-mode Gaussian state can be reconstructed (up to a reflection symmetry if the state is displaced and squeezed) by including a single loss-sensitive intensity measurement yielding the mean photon number or including simple click statistics in addition to $g^{(2)}$ and $g^{(3)}$ measurements. This discussion is extended to the multimode case in Sec.~\ref{sec:Multimode}. Finally, Sec.~\ref{sec:Summary} summarizes our findings.

\section{On general Gaussian states}\label{sec:GaussianStates}
In the following section, we introduce the essential mathematical concepts to describe Gaussian states providing a general framework on this class of states which has been studied extensively \cite{yeoman_twomode_1993,yuen_twophoton_1976,schumaker_new_1985,schumaker_quantum_1986,zhang_coherent_1990,ma_multimode_1990,simon_quantumnoise_1994,braunstein_quantum_2005,ferraro_gaussian_2005,vallone_means_2019,brask_gaussian_2022,cardin_photonnumber_2023}. Let us first introduce some notational conventions. In order to describe quantum states and in particular Gaussian states of multimode bosonic systems, we use standard creation and annihilation operators and employ a vector notation for simplicity: $\veca = (\hat{a}_1, \cdots,\hat{a}_M)^\top$ and $\veca^\dagger = (\hat{a}^\dagger_1, \cdots, \hat{a}^\dagger_M)$, where conjugation affects both the modal and vector structure, whereas transposition only alters the vector structure. $M$ specifies the number of modes. These operators obey the well known and defining commutation relation
\begin{equation}\label{eq.1}
    \big[\hat{a}_i,\hat{a}^\dagger_j\big] = \delta_{ij}.
\end{equation}

Using these operators, the fundamental Gaussian unitaries can be constructed:
\begin{subequations}\label{eq.2}
    \begin{align}
        \mathrm{Displacement}&:   &\hat{D}(\vec{\alpha})      &= \exp{\veca^\dagger\cdot\vec\alpha - \vec\alpha^\dagger\cdot\veca}\\ 
        \mathrm{Squeezing}&:     &\hat{S}(\boldsymbol{z})    &= \exp{\veca^\top\frac{\boldsymbol{z}^\dagger}{2}\veca -            \veca^\dagger\frac{\boldsymbol{z}}{2}\veca^{\dagger\top}}\\
        \mathrm{Rotation}&:      &\hat{R}(\boldsymbol{\phi}) &= \exp{\im\veca^\dagger\boldsymbol\phi\veca},
    \end{align}
\end{subequations}
which act as the fundamental operators under which Gaussianity is preserved. Here, $\vec\alpha$ is a complex-valued vector, $\boldsymbol{z}$ is a complex-valued symmetric matrix, and $\boldsymbol\phi$ is a hermitian matrix. Applying the commutation relations defined in Eq.~\eqref{eq.1} and expanding the exponentials, we can analyze the transformations of the bosonic ladder operators with respect to the Gaussian unitaries,
\begin{subequations}\label{eq.3}
    \begin{align}
        \hat{D}^\dagger(\vec\alpha) \hat{\vec{a}} \hat{D}(\vec\alpha)               &= \hat{\vec{a}} + \vec\alpha,\\
        \hat{R}^\dagger(\boldsymbol{\phi})\hat{\vec{a}}\hat{R}(\boldsymbol{\phi})   &= \E^{\im\boldsymbol\phi}\hat{\vec{a}},\\
        \hat{S}^\dagger(\boldsymbol{z})\hat{\vec{a}}\hat{S}(\boldsymbol{z})         &= \cosh(\boldsymbol{r})\hat{\vec{a}} - \sinh(\boldsymbol{r})\E^{\im\boldsymbol{\theta}}\hat{\vec{a}}^{\dagger\top},
    \end{align}
\end{subequations}
where $\boldsymbol{z} = \boldsymbol{r}\E^{\im\boldsymbol\theta}$ is the left polar decomposition of $\boldsymbol{z}$. The order in which these operations are applied to a state can be chosen arbitrarily, if the parameters are adjusted accordingly, since they can be rearranged using the established relations in Eqs.~\eqref{eq.3},
\begin{subequations}\label{eq.4}
    \begin{align}
        \hat{R}^\dagger(\boldsymbol{\phi}) \hat{D}(\vec\alpha)\hat{R}(\boldsymbol\phi) 
        &= \hat{D}\big(\E^{-\im\boldsymbol\phi}\vec\alpha\big),\\
        \hat{R}^\dagger(\boldsymbol{\phi}) \hat{S}(\boldsymbol{z})\hat{R}(\boldsymbol\phi) 
        &= \hat{S}\big(\E^{-\im\boldsymbol\phi}\boldsymbol{z}\E^{-\im\boldsymbol\phi^\top}\big),\\
        \hat{S}^\dagger(\boldsymbol{z})\hat{D}(\vec\alpha)\hat{S}(\boldsymbol{z})
        &= \hat{D}\big(\cosh(\boldsymbol{r})\vec\alpha + \sinh(\boldsymbol{r})\E^{\im\boldsymbol\theta}\vec\alpha^*\big).
    \end{align}
\end{subequations}

A general Gaussian unitary $\hat{U}$ is then constructed using these three operations. Its effect on a state can be characterized in various ways, including examining its influence on the creation and annihilation operators. This influence can be represented by a Bogoliubov transformation. It assigns an affine vector transformation to the vector operator $\hat{b}_\mu = (\hat{a}_1,\cdots,\hat{a}_M, \hat{a}^\dagger_1,\cdots \hat{a}^\dagger_M)^\top$, reflecting the action of the fundamental unitaries:
\begin{equation}\label{eq.5}
        \hat{b}'_\mu = \hat{U}^\dagger \hat{b}_\mu \hat{U} = L_\mu^{\phantom{\mu}\nu}\hat{b}_\nu + A_\mu ,
\end{equation}
with a $2M\times 2M$ matrix $\boldsymbol{L}$ encoding squeezings and rotations and a $2M\times1$ vector $\vec{A}$ representing displacements. In particular, if $\hat{U} = \hat{D}(\vec\alpha)\hat{S}(\boldsymbol{z})\hat{R}(\boldsymbol{\phi})$: 
\begin{equation}\label{eq.6}
    \begin{aligned}
       \vec{A} &=\begin{pmatrix}
            \vec{\alpha}\\
            \vec{\alpha}^*
        \end{pmatrix}, 
        & \boldsymbol{L} &=
        \begin{pmatrix}
           \boldsymbol{E} &\boldsymbol{F}\\
           \boldsymbol{F}^* & \boldsymbol{E}^*
        \end{pmatrix}, \\
        \boldsymbol{E} &= \cosh(\boldsymbol{r})\E^{\im\boldsymbol{\phi}}, & \boldsymbol{F}&=  -\sinh(\boldsymbol{r})\E^{\im\boldsymbol{\theta}}\E^{-\im\boldsymbol{\phi}^\top}, 
    \end{aligned}
\end{equation}
and the Bogoliubov transformation immediately provides the combined relations of Eqs.~\eqref{eq.3} by considering the first $M$ elements after applying Eq.~\eqref{eq.5}.

In order to generate the most general Gaussian state, the fundamental unitaries may act on an uncorrelated multimode thermal state 
\begin{subequations}\label{eq:thermalState}
    \begin{align}
        \hat\rho_\mathrm{th,M} &= \bigotimes_{k=1}^{M}\hat\rho_\mathrm{th}^{(k)}(N_k), \\
        \hat\rho^{(k)}_\mathrm {th}(N_k) &= \sum_{n=0}^\infty \frac{N_k^n}{(N_k + 1)^{n+1}}\ket{n}_k\!\bra{n}_k, 
    \end{align}
\end{subequations}
where each single mode thermal state $\hat\rho^{(k)}_\mathrm{th}$ with expected photon number $N_k = \mathrm{tr} \big(\hat\rho^{(k)}_\mathrm {th} \hat{a}_k^{\dagger}\hat{a}_k \big)$ is written in Fock basis. Combining this multimode thermal state with the fundamental unitaries results in the most general multimode Gaussian state which reads:
\begin{equation}\label{eq:multiGaussian}
\hat\rho_\mathrm{g}  = \hat{D}(\vec\alpha)\hat{S}(\boldsymbol{z})\hat{R}(\boldsymbol{\phi}) \,\hat\rho_\mathrm{th,M}\, \hat{R}^\dagger(\boldsymbol{\phi})\hat{S}^\dagger(\boldsymbol{z})\hat{D}^\dagger(\vec\alpha).
\end{equation}
Note that for $N_k = 0$ for all  $k \in \{1,\cdots,M \}$, the multimode thermal state reduces to the vacuum. Thus, all pure Gaussian states are also included as they can be generated by acting with the fundamental Gaussian operations on the vacuum state.

Finally, counting the degrees of freedom yields
\begin{equation}\label{eq.9}
    \centering
    \begin{tabular}{c|c}
        $\hat{\rho}^{M}_\mathrm{th}$                 & $M$       \\
        $\hat{D}(\vec\alpha)$                        & $2M$      \\
        $\hat{R}(\boldsymbol{\phi})$                 & $M^2-M$   \\
        $\hat{S}(\boldsymbol{z})$                    & $M^2 + M$ \\ \hline
        $\hat{\rho}_\mathrm{g}$                               & $2M^2 + 3M$
    \end{tabular}
\end{equation}
where $\hat{R}$ only contributes $M^2 -M$ degrees of freedom since the diagonal elements of $\boldsymbol{\phi}$ are paired with operators $\im \hat{a}_i^\dagger \hat{a}_i$ such that the respective transformations merely correspond to an unobservable global phase term that does not alter the density operator given in Eq.~\eqref{eq:thermalState}.

\section{Correlation functions and moment decompositions}\label{sec:CorrelationFunctions}
For arbitrary quantum states, the second- and third-order intensity correlation functions at zero time delay are defined in terms of operator moments \cite{glauber_quantum_1963,loudon_the_1983,vogel_quantum_2006,laiho_measuring_2022},
\begin{equation}\label{eq.10}
 g^{(2)}_{ij} = \frac{\langle \hat{a}_i^\dagger \hat{a}^\dagger_j \hat{a}_i\hat{a}_j\rangle}{\langle \hat{a}^\dagger_i\hat{a}_i\rangle\langle\hat{a}^\dagger_j\hat{a}_j\rangle}, \quad
  g^{(3)}_{ijk} = \frac{\langle \hat{a}_i^\dagger \hat{a}^\dagger_j \hat{a}^\dagger_k\hat{a}_i\hat{a}_j \hat{a}_k\rangle}{\langle \hat{a}^\dagger_i\hat{a}_i\rangle\langle\hat{a}^\dagger_j\hat{a}_j\rangle\langle\hat{a}_k^\dagger\hat{a}_k\rangle}.
\end{equation}
However, this notation can be cumbersome for intricate calculations. In order to simplify the notation of the higher-order moments of the ladder operators, we introduce the shorthand notations $\mo{\us i}= \langle \hat{a}^\dagger_i\rangle$, $\mo{i}= \langle \hat{a}_i\rangle$, and $\mo{\us i j}= \langle \hat{a}^\dagger_i \hat{a}_j\rangle$, which is generalized to arbitrary operator moments as
\begin{equation}\label{eq.11}
        \mo{\us {i_1}\cdots \us {i_n} j_1\cdots j_m} = \langle\hat{a}^\dagger_{i_1}\cdots\hat{a}^\dagger_{i_n}\hat{a}_{j_1}\cdots \hat{a}_{j_m} \rangle,
\end{equation}
where underlined indices refer to the creation-operator indices while normal indices denote an index attached to an annihilation operator. Note that the second-order moment $\mo{\us i i} = \langle \hat{a}^\dagger_i\hat{a}_i\rangle = \bar{n}_i$ corresponds to the expectation value of the number operator in the $i$-th mode. Further, $\mo{\us i j}$ is the unnormalized first-order coherence function between mode $i$ and $j$ encoding field correlations. Similarly, we identify unnormalized second- and third-order correlation functions with the operator moments $\mo{\us i\us j i j}$ and $\mo{\us i\us j \us k ijk}$, respectively, thus
\begin{align}\label{eq.12}
        g^{(1)}_{ij}      &= \frac{\mo{\us i j}}{\sqrt{\bar{n}_i \bar{n}_j}},&
        g^{(2)}_{ij}      &= \frac{\mo{\us i\us j i j}}{\bar{n}_i \bar{n}_j},&
        g^{(3)}_{ijk}     &= \frac{\mo{\us i\us j \us k ijk}}{\bar{n}_i\bar{n}_j\bar{n}_k}.
\end{align}

As we focus on Gaussian states, any expectation value of ladder operators can be written in terms of first- and second-order expectation values. This can be traced back to the fact that any Gaussian state is uniquely defined by its first- and second-order moments of quadrature operators which are connected to the ladder operators by a simple unitary transformation implying a basis change in phase space between real-valued and complex-valued coordinates. For instance, 
\begin{equation}\label{eq.13}
    \begin{aligned}
        \mo{\mu\nu\kappa\lambda} &= \mo{\mu\nu}\mo{\kappa\lambda} + \mo{\mu\kappa}\mo{\nu\lambda} + \mo{\mu\lambda}\mo{\nu\kappa}\\
        &\quad -2\,\mo{\mu}\mo{\nu}\mo{\kappa}\mo{\lambda},
    \end{aligned}
\end{equation}
where Greek indices refer to either underlined or normal Latin indices. A scheme of this procedure has been established in Ref.~\cite{vallone_means_2019}. In order to break down the third-order correlation function in the same way, an equation similar to Eq.~\eqref{eq.13} needs to be established for sixth-order moments $\mo{\us i \us j\us k i j k}$. We provide a full derivation of such a decomposition for a general Gaussian state in App.~\ref{app.A}. In the following, we will motivate the final result on structural ground based on the simpler decomposition of the fourth-order moment in Eq.~\eqref{eq.13}.

The set of indices $\{\mu,\nu,\kappa,\lambda\}$ from the fourth-order moment in Eq.~\eqref{eq.13} can be used to construct the products of second-order moments on the right-hand side by considering all possible partitions of this set into $2$ subsets of size $2$. Of course, the ordering needs to be respected between indices. This partitioning can be generalized to higher-order moments. 
A set $\omega$ of cardinality $ \#\omega = x \cdot y$ can be partitioned into $x$ disjoint subsets of size $y$. We denote by $\mathcal{P}^x_y(\omega)$ the set of all partitions $s$ fulfilling this requirement. For example, in case of Eq.~\eqref{eq.13}, the set of indices $\{\mu,\nu,\kappa,\lambda\}$ has a cardinality of $4$, with $x = 2$ and $y = 2$ such that $\mathcal{P}_2^2(\{\mu,\nu,\kappa,\lambda\}) = \Big\{\big\{\{\mu,\nu\},\{\kappa,\lambda\} \big\},\big\{\{\mu,\kappa\},\{\nu,\lambda\} \big\},\big\{\{\mu,\lambda\},\{\nu,\kappa\} \big\}\Big\}$. The partitions then correspond to the products in Eq.~\eqref{eq.13} such that each moment is represented by a subset within a given partition $s$. We can then refer to the specific indices by indexing the partitions $s$: $s_{ij}$ indexes the $j$th element in the $i$th subset of $s$. To exemplify, consider $s = \big\{\{\mu,\nu\},\{\kappa,\lambda\}\big\}$ where $s_{11} = \mu$, $s_{12} = \nu$, $s_{21} = \kappa$, and $s_{22} = \lambda$.

Nonetheless, this strategy will not yet suffice to fully decompose an arbitrary sixth-order moment for Gaussian states, as we have not considered contributions containing products involving first-order moments. However, it is in fact sufficient to perform the decomposition for thermal states whose first-order moments vanish. These can then in turn be transformed to moments of arbitrary Gaussian states using Bogoliubov transformations as outlined in App.~\ref{app.A}.

To characterize the remaining terms that will arise in the decomposition, we need one more ingredient. Let us define $\mathcal{P}_{v,w}(\omega)$, which represents the set of all bipartitions $\{\psi, \chi\}$ of $\omega$ such that $\#\omega = v+w$ with $\#\psi = v$ and $\#\chi=w$. Indices in $\psi_i$ and $\chi_i$ refer to the elements of the respective sets. For a more in-depth explanation of the general formalism using examples see App.~\ref{app.A} as well.

With these ingredients, the decomposition formula for the sixth-order moment can now be written as
\begin{equation}\label{eq.15}
    \begin{aligned}
        \mo{\mu\nu\kappa\lambda\rho\sigma}
        &= \sum_{s\in\mathcal{P}^3_2(\omega)}\prod_{i=1}^3 \mo{s_{i1} s_{i2}} \\
        &\qquad-2 \!\!\!\!\sum_{\substack{\{\psi,\chi\} \\ \in\mathcal{P}_{4,2}(\omega)}}\!\!\! \mo{\chi_1 \chi_2} \mo{\psi_1}\mo{\psi_2}\mo{\psi_3}\mo{\psi_4}\\
        &\qquad+ 16\mo{\mu}\mo{\nu}\mo{\kappa}\mo{\lambda}\mo{\rho}\mo{\sigma},
    \end{aligned}
\end{equation}
where the summations run over the aforementioned partitions of the set of indices $\omega = \big\{\mu,\nu,\kappa,\lambda,\rho,\sigma \big\}$ implying that each sum contains 15 terms. In general, similar decomposition formulas for arbitrarily higher correlation functions can be constructed using the scheme outlined in this section and in App.~\ref{app.A}, albeit at the cost of an increasing number of contributing terms.

Structurally, our approach constitutes a generalized application of Isserlis’ theorem, as we do not restrict to zero-mean Gaussian states \cite{isserlis_formula_1918}. Related decompositions have been considered in the literature, predominantly in the context of photon number cumulants rather than correlation functions, and it has been shown that higher-order moments can be expressed in terms of loop hafnians \cite{cardin_photonnumber_2023}. Although compact, this representation is not ideally suited for term-wise manipulations. The decomposition employed here is, nevertheless, informationally equivalent.

In order to determine a dependence between second- and third-order correlation functions using the decomposition formulae, we first consider $g^{(2)}$ in terms of the introduced formalism. Using Eq.~\eqref{eq.13} to decompose the fourth-order moment in $g^{(2)}$ and setting the arbitrary greek indices to $\mu = \us i$, $\nu = \us j$, $\kappa = i$, $\lambda = j$, as well as normalizing the fourth-order moment via the mean photon numbers of the respective modes, we find 
\begin{align}
      g^{(2)}_{ij} &= 1 + \frac{\mo{\us i \us j}\mo{ij} + \mo{\us i j}\mo{\us j i}-2|\mo{i}|^2|\mo{j}|^2}{\bar{n}_i \bar{n}_j} \notag\\
      &= 1 + |g_{ij}^{(1)}|^2 + \frac{|\cov{ij}+\alpha_i\alpha_j|^2 - 2|\alpha_i|^2|\alpha_j|^2}{\bar{n}_i \bar{n}_j},
      \label{eq:g2Decomposition}
\end{align}
where we also used the fact that for Gaussian states $\mo{i}\equiv\langle \hat{a}_i\rangle = \alpha_i$ within our conventions. Further, the second-order operator moment $\mo{ij}$ is rewritten in terms of its centralized from, i.e., the covariance of the annihilation operators, $\cov{ij} \equiv \langle \hat{a}_i \hat{a}_j\rangle - \langle\hat{a}_i\rangle\langle\hat{a}_j\rangle = \mo{ij} - \alpha_i \alpha_j$.

Equivalently, we are able to decompose the third-order correlation function $g^{(3)}$ into at most second-order expectation values given by $\alpha_i$, $\bar{n}_i$, $g^{(1)}_{ij}$, and $\cov{ij}$ with the aid of Eq.~\eqref{eq.15},
\begin{widetext}
    \begin{align}
    g^{(3)}_{ijk}
    &= 1 + |g_{ij}^{(1)}|^2 + |g_{jk}^{(1)}|^2 + |g_{ik}^{(1)}|^2 + 2\mathrm{Re}[g^{(1)}_{ij}g^{(1)}_{jk}g^{(1)}_{ki}]
    + \frac{|\cov{ij}|^2 - |\alpha_i \alpha_j|^2}{\bar{n}_i \bar{n}_j}
    + \frac{|\cov{jk}|^2 - |\alpha_j \alpha_k|^2}{\bar{n}_j \bar{n}_k}
    + \frac{|\cov{ik}|^2 - |\alpha_i \alpha_k|^2}{\bar{n}_i \bar{n}_k} 
    \notag \\
    &\quad  + \bigg(2 - 4\frac{|\alpha_k|^2}{\bar{n}_k}\bigg) \frac{ \mathrm{Re}[\mathrm{cov}_{ij}\alpha_i^*\alpha_j^*]}{\bar{n}_i \bar{n}_j}
     + \bigg(2 - 4\frac{|\alpha_j|^2}{\bar{n}_j}\bigg) \frac{\mathrm{Re}[\mathrm{cov}_{ik}\alpha_i^*\alpha_k^*]}{\bar{n}_i \bar{n}_k}
     + \bigg(2 - 4\frac{|\alpha_i|^2}{\bar{n}_i}\bigg) \frac{\mathrm{Re}[\mathrm{cov}_{jk}\alpha_j^*\alpha_k^*]}{\bar{n}_j \bar{n}_k}
     + 4 \frac{|\alpha_i \alpha_j \alpha_k|^2}{\bar{n}_i \bar{n}_j \bar{n}_k}  
     \notag \\
     &\quad + 2 \frac{\mathrm{Re}\Big[g_{ij}^{(1)} \cov{jk}^*\cov{ik}\Big]}{\bar{n}_k \sqrt{\bar{n}_i \bar{n}_j}}
     + 2 \frac{\mathrm{Re}\Big[g_{jk}^{(1)} \cov{ik}^*\cov{ij}\Big]}{\bar{n}_i \sqrt{\bar{n}_j \bar{n}_k}}
     + 2 \frac{\mathrm{Re}\Big[g_{ki}^{(1)} \cov{ij}^*\cov{jk}\Big]}{\bar{n}_j \sqrt{\bar{n}_i \bar{n}_k}}
     - 2 \frac{|\alpha_k|^2}{\bar{n}_k} \frac{\mathrm{Re}\Big[g_{ij}^{(1)} \alpha_i\alpha_j^*\Big]}{\sqrt{\bar{n}_i \bar{n}_j}}
     \notag \\
     &\quad - 2 \frac{|\alpha_i|^2}{\bar{n}_i} \frac{\mathrm{Re}\Big[g_{jk}^{(1)} \alpha_j\alpha_k^*\Big]}{\sqrt{\bar{n}_j \bar{n}_k}}
     - 2 \frac{|\alpha_j|^2}{\bar{n}_j} \frac{\mathrm{Re}\Big[g_{ki}^{(1)} \alpha_k\alpha_i^*\Big]}{\sqrt{\bar{n}_i \bar{n}_k}}
     + 2\frac{ \mathrm{Re}\big[g_{ij}^{(1)}\alpha_i\alpha_k\cov{jk}^* \big] + \mathrm{Re}\big[g_{ij}^{(1)}\alpha_j^*\alpha_k^*\cov{ik}\big]}{\bar{n}_k \sqrt{\bar{n}_i \bar{n}_j}} 
     \notag \\
     &\quad + 2\frac{ \mathrm{Re}\big[g_{jk}^{(1)}\alpha_i\alpha_j\cov{ik}^* \big] + \mathrm{Re}\big[g_{jk}^{(1)}\alpha_i^*\alpha_k^*\cov{ij}\big]}{\bar{n}_i \sqrt{\bar{n}_j \bar{n}_k}} 
     + 2\frac{ \mathrm{Re}\big[g_{ki}^{(1)}\alpha_j\alpha_k\cov{ij}^* \big] + \mathrm{Re}\big[g_{ki}^{(1)}\alpha_i^*\alpha_j^*\cov{jk}\big]}{\bar{n}_j \sqrt{\bar{n}_i \bar{n}_k}}. 
     \label{eq:g3Decomposition}
    \end{align}
\end{widetext}
Based on this decomposition, we can formally derive a relation between $g^{(3)}$ and $g^{(2)}$ in a similar fashion to what has been presented in the literature for single-mode squeezed vacuum \cite{wakui_ultrabroadband_2014}. For this particular case, one obtains $g^{(2)} = 3 + \frac{1}{\sinh^2r}$ and $g^{(3)} = 15 + \frac{9}{\sinh^2r}$. As both correlation functions only depend on the squeezing magnitude $r$, one finds a unique dependence of $g^{(3)}$ on $g^{(2)}$, $g^{(3)} = 9g^{(2)}-12$. For more general Gaussian states, e.g., displaced squeezed vacua or multimode scenarios, however, the correlation functions involve additional parameters. Therefore, several equivalent relations may exist, as we have the freedom to choose which parameters of the state to express in terms of $g^{(2)}$.

\section{Dependence of correlations and state reconstruction for Single-mode Gaussian States}\label{sec:SingleMode}
First, we restrict the discussion to the single-mode case but allow for a general Gaussian structure. Then, the correlation functions read 
\begin{align}
 g^{(2)} &= 2 + \frac{|\cov{}|^2 + 2 \mathrm{Re}\big( \alpha^2 \cov{}^* \big) - |\alpha|^4}{\bar{n}^2},
 \label{eq:g2SingleMode}
 \\
 g^{(3)} &= 6 + 9\frac{|\cov{}|^2 + 2 \mathrm{Re}\big( \alpha^2 \cov{}^* \big) - |\alpha|^4}{\bar{n}^2} \notag \\
 &\quad + \frac{4|\alpha|^6 - 12|\alpha|^2 \mathrm{Re}\big( \alpha^2 \cov{}^* \big)}{\bar{n}^3},
 \label{eq:g3SingleMode}
\end{align}
where we dropped the superfluous index notation (single-mode) and $\cov{} = -\E^{\im \theta}(2N+1) \sinh r \cosh r$ becomes the variance of $\hat{a}$ being a function of the squeezing strength $r$, the squeezing phase $\theta$, and the thermal mean occupation number $N$. Further, the mean photon number is given by ${\bar{n} =  |\alpha|^2 + \sinh^2r + N(\cosh^2r+\sinh^2r)}$.  
As expected, the intensity correlations depend on four of the five parameters characterizing a single-mode Gaussian state: $|\alpha|$, $r$, $N$, as well as the relative phase between squeezing and displacement encoded in $\mathrm{Re}\big( \alpha^2 \cov{}^* \big)$. The remaining fifth parameter is an overall rotation in phase space, which only acquires meaning relative to an external phase reference and has no observable consequences for photon-statistics measurements; it just reflects the arbitrariness of the phase-space coordinate system.

From Eqs.~\eqref{eq:g2SingleMode} and \eqref{eq:g3SingleMode}, it can be deduced that a unique $g^{(2)}$-$g^{(3)}$ relation can be established for any single-mode Gaussian state with either vanishing displacement or vanishing squeezing. In these cases, both correlation functions only depend either on the ratio $|\cov{}|/\bar{n}$ ($\alpha=0$) or $|\alpha|^2/\bar{n}$ ($\cov{}=0$). Thus, we obtain the unique relation 
\begin{align}
 g^{(3)} = 9g^{(2)}-12
 \label{eq:g3-g2singleSV}
\end{align}
for any single-mode Gaussian state without displacement, i.e., a squeezed vacuum or squeezed thermal state, and 
\begin{align}
 g^{(3)} = 9g^{(2)}-12 + 4(2-g^{(2)})^{\frac{3}{2}}
  \label{eq:g3-g2singleDT}
\end{align}
for any single-mode Gaussian state without squeezing (displaced vacuum or displaced thermal state).

For the case of a single-mode Gaussian state with nonvanishing displacement and squeezing, we have the freedom to rewrite either of the four quantities $\bar{n}$, $|\alpha|$, $|\cov{}|$, or $\mathrm{Re}\big( \alpha^2 \cov{}^* \big)$ in terms of $g^{(2)}$ and insert this expression into $g^{(3)}$. A convenient choice would be a quantity that is sophisticate to measure or easy to manipulate on practical grounds based on the available resources. For illustrative purposes, we choose $\mathrm{Re}\big( \alpha^2 \cov{}^* \big)$ such that a potential $g^{(2)}$-$g^{(3)}$ relation for the general single-mode case reads 
\begin{align}
 g^{(3)}_{} &= \Big[9-6\frac{|\alpha|^2}{\bar{n}}\Big]g^{(2)} + 2\frac{|\alpha|^2}{\bar{n}}\Big[ 6 + \frac{3 |\cov{}|^2 - |\alpha|^4}{\bar{n}^2}\Big] - 12.
 \label{eq:g3-g2singleDST}
\end{align}

With the aid of Eqs.~\eqref{eq:g3-g2singleSV}-\eqref{eq:g3-g2singleDST}, we can now perform hypothesis testing for any single-mode Gaussian state. If the state of interest does not satisfy any of these relations, it can be concluded with certainty that it is multimode or non-Gaussian. By contrast, fulfilling one of the three relations provides at least evidence for a single-mode Gaussian state. Certainly non-Gaussian counterexamples exist, e.g., the Fock-diagonal mixture $ \frac{5}{8}\ket{0}\!\!\bra{0}+\frac{3}{8}\op{2}$ yields $g^{(2)}=\frac{4}{3}$ and $g^{(3)}=0$ such that Eq.~\eqref{eq:g3-g2singleSV} would be satisfied. 
Note, however, that any single-mode squeezed thermal Gaussian state fulfilling Eq.~\eqref{eq:g3-g2singleSV} must in addition satisfy $g^{(2)}\geq 2$ and $g^{(3)}\geq 6$ (see Eqs.~\eqref{eq:g2SingleMode} and \eqref{eq:g3SingleMode} for $\alpha=0$), so the above mixture, with $g^{(2)}=\frac{4}{3}$ and $g^{(3)}=0$, can still be excluded as Gaussian. Nevertheless, genuine families of non-Gaussian states exist that perfectly mimic a Gaussian point in the $(g^{(2)},g^{(3)})$ plane. For instance, for any $0<c \leq \frac{1}{3}$ the Fock-diagonal state $(1-c+c^2-c^3) \op{0} + (c-2c^2+3c^3) \op{1} + (c^2-3c^3) \op{2} + c^3 \op{3}$ yields fixed $g^{(2)} = 2$ and $g^{(3)} = 6$, identical to a single-mode thermal state, while being manifestly non-Gaussian. 
We emphasize that this ambiguity is not specific to our $g^{(2)}$-$g^{(3)}$-based analysis but is a generic feature of any characterization scheme based on finitely many measured observables. A finite data set imposes only finitely many constraints on the density operator. Thus, the set of compatible states typically contains both Gaussian and non-Gaussian states. For example, the same caveat applies to homodyne detection with a finite grid of quadrature phases and finite resolution, where non-Gaussian states can in principle reproduce all measured quadrature data while differing in unsampled regions of phase space.

\begin{figure}[t]
    \centering
    \includegraphics[width=1\columnwidth]{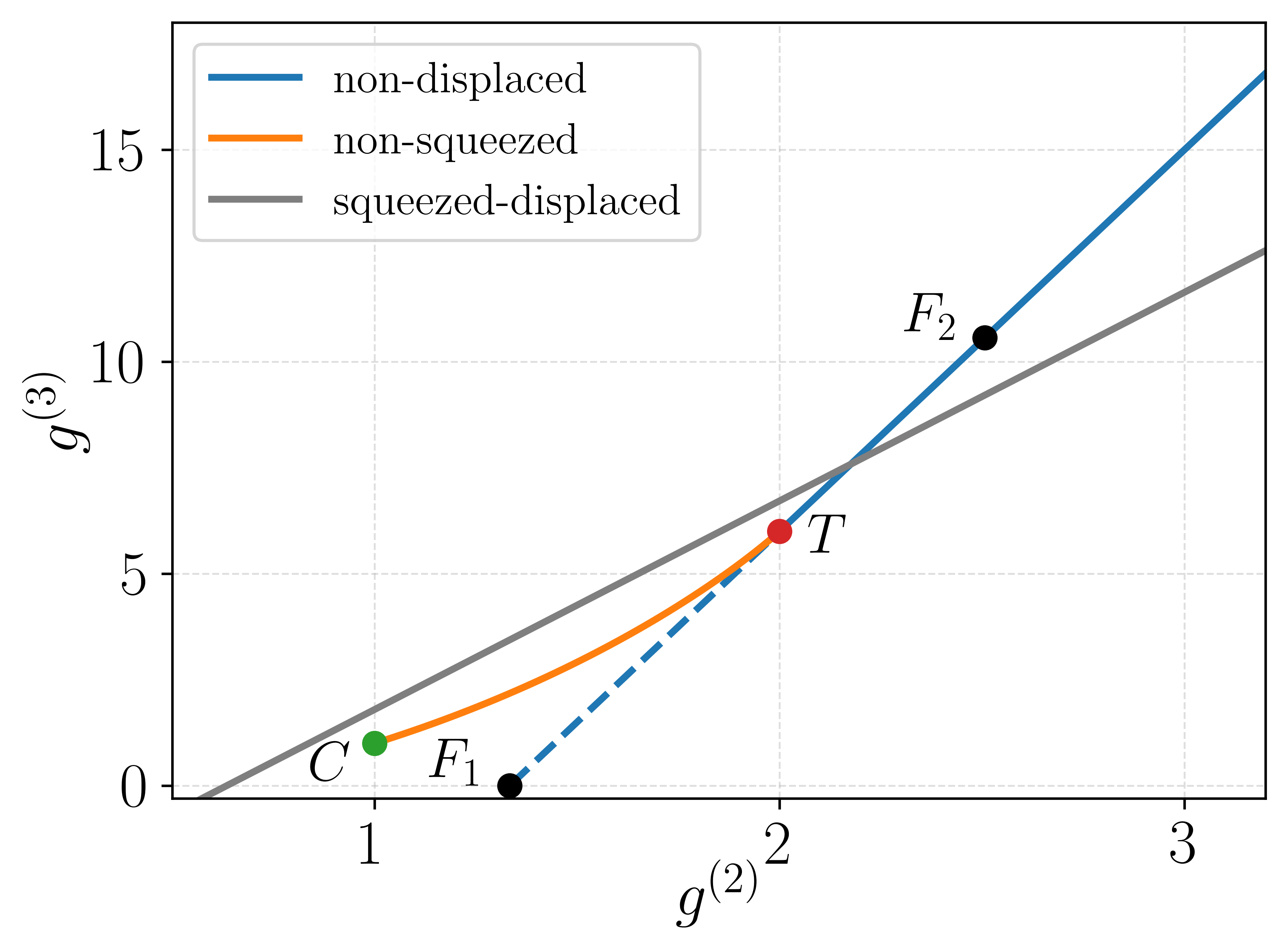}
    \caption{Functional dependence $g^{(3)}(g^{(2)})$ for different subclasses of the Gaussian sector. Coherent states (point $C$, green) and thermal states (point $T$, red) are shown together with the relations of non-squeezed ($1 \leq g^{(2)} \leq 2$) and non-displaced ($2 \leq g^{(2)}$) states. Note that the domains of these classes are bounded, cf. Eqs.~\eqref{eq:g2SingleMode} for $\cov{}=0$ (note that $\bar{n} = |\alpha|^{2}+N$ in this case) and $\alpha=0$, respectively. A squeezed-displaced example following Eq.~\eqref{eq:g3-g2singleDST} with $N = 0.5$, $\alpha = 5$, and $r = 1.6$ is plotted in gray. The two black points correspond to the Fock-state mixtures $F_1 = \tfrac{5}{8}\op{0} + \tfrac{3}{8}\op{2}$ and $F_2 \approx 0.9\op{2} + 0.1\op{18}$, both lying exactly on the linear relation for non-displaced Gaussian states, Eq.~\eqref{eq:g3-g2singleSV}. The dashed continuation of the blue curve illustrates that this functional dependence may also be fulfilled by certain non-Gaussian states outside the allowed domain for Gaussian states (point $F_{1}$).}
    \label{fig:g3g2SMplot}
\end{figure}

Nevertheless, evidence for a single-mode Gaussian state can be strengthened if parameter scans are possible. In case independent source adjustments can be done that generate states within the same class of Gaussian states but with altered $g^{(2)}$ and $g^{(3)}$ values, we are able to test the functional dependency $g^{(3)}(g^{(2)})$ over some para\-meter range. For instance, let us consider a source that generates single-mode squeezed vacuum. If one is able to experimentally manipulate the squeezing strength of the source, we modify $g^{(2)}$ as well as $g^{(3)}$ but Eq.~\eqref{eq:g3-g2singleSV} continues to hold. Such an approach would substantially increase support for the corresponding single-mode Gaussian hypothesis. Additionally, we obtain the information that the Gaussian state is not displaced, not squeezed, or exhibits displacement and squeezing, depending on which of the three relations is satisfied. In Fig.~\ref{fig:g3g2SMplot} we visualized these relations in the $g^{(2)}$-$g^{(3)}$ plane.

We can further refine the hypothesis test by performing an additional intensity measurement yielding the mean photon number $\bar{n}$. This additional information allows us, in combination with the correlation functions, to eventually determine most of the parameters specifying a single-mode Gaussian state. However, we would like to emphasize that there is one crucial difference between the $g^{(n)}$ observables and $\bar{n}$. While the normalized correlation functions $g^{(n)}$ are invariant quantities under linear loss, the measurement of the mean photon number is affected by the detector efficiency and losses introduced by all optical elements in the path from the source to the detector. Thus, this measurement needs careful calibration. 
Albeit including a loss-sensitive quantity complicates the data analysis, it turns out that such an observable is crucial to determine specific state properties that are not accessible via loss-invariant observables.

The reason for this can be traced back to the fact that Gaussian states remain Gaussian under linear loss. Trivial examples would be classical states such as coherent or thermal states which remain coherent or thermal under loss, respectively. Indeed, all coherent states will result in $g^{(2)} = g^{(3)} = g^{(4)} = \cdots = 1$. Therefore, we can only conclude the presence of a coherent state from intensity correlation measurements but can never infer the displacement. However, additionally measuring the mean photon number $\bar{n}$ allows us to derive the displacement magnitude. Similarly, for thermal states, we have $g^{(n)} =n!$ but measuring $\bar{n}$ allows to infer the thermal mean occupation $N$. This discussion becomes more involved for more complex states. Consider a single-mode Gaussian state exhibiting squeezing but no displacement. Such a state satisfies Eq.~\eqref{eq:g3-g2singleSV} which is a well known relation for single-mode squeezed vacuum states \cite{christ_probing_2011,wakui_ultrabroadband_2014}. Thus it would be tempting to combine the $g^{(2)}$ measurement result of this state with the fact that $g^{(2)} = 3 + \frac{1}{\sinh^2r}$ for squeezed vacuum to extract the squeezing parameter $r$. However, our analysis clearly reveals that any squeezed thermal state also satisfies Eq.~\eqref{eq:g3-g2singleSV}. Therefore, extracting $r$ from the correlation measurement alone can be flawed if the actual state is a squeezed thermal state rather than a pure squeezed vacuum state. Measuring $\bar{n}$ and verifying whether the relation $g^{(2)} = 3 + \frac{1}{\bar{n}}$ holds provides a criterion for distinguishing between pure and mixed squeezed single-mode states. In fact, we demonstrate in the following that we can strongly constrain any single-mode Gaussian state by combining the $g^{(2)}$-$g^{(3)}$ relations with a loss sensitive observable as the mean photon number.

Apart from using $\bar{n}$, also other loss-sensitive observables can be used to specify concrete state parameters on quantitative grounds. A convenient example would be the overlap of the Gaussian state with the vacuum state which would be a natural choice if the correlation functions are measured via click detectors in a low-flux scenario.

\subsection{Non-displaced single-mode Gaussian states}
For vanishing displacement (or squeezing), the hypothesis test for single-mode Gaussian states is straightforward. Measuring $g^{(2)}$ and $g^{(3)}$ is already sufficient as $g^{(3)}$ only depends on $g^{(2)}$ for those cases. 
If a state fulfills Eq.~\eqref{eq:g3-g2singleSV}, we have evidence for a zero-mean, single-mode Gaussian state. 
Having ${g^{(2)} = 2 + \frac{|\cov{}|^2}{\bar{n}^2}}$ with $|\cov{}|^2 = (2N+1)^2\cosh^2r\, \sinh^2r$ and ${\bar{n} = \sinh^2r + N(\cosh^2r+\sinh^2r)}$ for the most general state of this class, we can specify two of the three parameters characterizing any pure or mixed single-mode squeezed Gaussian state, 
\begin{align}
 \sinh^2 r &= \frac{1}{2} \frac{2\bar{n}+1}{\sqrt{(2\bar{n}+1)^2 - 4\bar{n}^2(g^{(2)}-2)}} - \frac{1}{2},
 \\
 N &= \frac{1}{2} \sqrt{(2\bar{n}+1)^2 - 4\bar{n}^2(g^{(2)}-2)} - \frac{1}{2}.
\end{align}
Note that the state is pure if $N=0$ implying $g^{(2)} = 3 + \frac{1}{\bar{n}}$ and thus $ \sinh^2 r = \bar{n}$ such that we recover the known result for single-mode squeezed vacuum states. Further, we can also deduce the presence of a thermal state as measuring $g^{(2)} = 2$ gives a vanishing squeezing magnitude.

The remaining quantity, the phase of the squeezing parameter $\theta$, merely describes a rotation in phase space and can only be determined via an external phase reference. If $\theta$ does not drift over relevant time scales, we have the gauge freedom to choose $\theta = 0$ without loss of generality by appropriately choosing the phase-space coordinate system. By contrast, if $\theta$ does drift, which implies that we have to effectively average over it, the squeezing will diminish and the underlying state need not necessarily remain Gaussian. If the state remains Gaussian, e.g., via a uniform average, we simply obtain altered squeezing amplitudes and mean thermal occupation numbers. If the state becomes non-Gaussian due to the drift, we are able to detect such a behavior via a violation of the derived $g^{(2)}$-$g^{(3)}$ relation given by Eq.~\eqref{eq:g3-g2singleSV}.

If click detectors are used to determine the correlation functions, it might be more convenient to use (no-)click statistics as a loss-sensitive observable instead of measuring the mean photon number. In such a scenario, we can extract the two parameters $r$ and $N$ from the $g^{(2)}$ value and the no-click probability $p_0 = \tr \big( \hat\rho_\mathrm{g} \op{0} \big)$ with $p_0^2 = \frac{1}{N^2 + (2N+1)\cosh^2 r}$. The thermal mean occupation number is obtained by finding the root of a fourth-order polynomial which is a straightforward numerical task,
\begin{align}
 0 &= N^4 + 2N^3 + \Big(g^{(2)}-1-\frac{2}{p_0^2}\Big)N^2 \\
 &\quad + \Big(g^{(2)}-2-\frac{2}{p_0^2}\Big)N + \Big(1-\frac{1}{p_0^2}\Big) \Big(g^{(2)}-2-\frac{1}{p_0^2}\Big). \notag
\end{align}
If $N$ is known, we are able to eventually obtain the squeezing amplitude via 
\begin{align}
 \sinh^2r = \frac{1-(N+1)^2p_0^2}{(2N+1)p_0^2}.
\end{align}

\subsection{Non-squeezed single-mode Gaussian states}
Similarly, we can determine the displacement amplitude and mean thermal occupation number if a state fulfills Eq.~\eqref{eq:g3-g2singleDT} providing evidence for a displaced thermal state. Then, 
\begin{align}
 |\alpha|^2 &= \bar{n} \sqrt{2-g^{(2)}}, \qquad 
 N = \bar{n} \Big(1-\sqrt{2-g^{(2)}} \Big).
\end{align}
Note that this result also includes the special case of a displaced vacuum state. If $g^{(2)} = 1$ is measured, we have $N=0$ and $|\alpha|^{2} = \bar{n}$. Nontrivial evidence can be obtained by checking if $g^{(3)} = 1$, as well as by scanning through different intensities of the state yielding the same results for any intensity correlations, $g^{(2)} = g^{(3)} = 1$. Analogously, we recover the properties of a thermal state, if we measure $g^{(2)} = 2$, implying $\alpha=0$ and $N=\bar{n}$, with a nontrivial check via $g^{(3)} = 6$ for all $N$. 

Using the vacuum overlap instead of an intensity measurement as a loss-sensitive observable, we can numerically infer $|\alpha|$ and $N$ from
\begin{align}
 p_0 = \frac{1}{N+1} \E^{-\frac{|\alpha|^2}{N+1}} \quad \text{and} \quad \sqrt{2-g^{(2)}} = \frac{|\alpha|^2}{|\alpha|^2 + N}.
\end{align}
Similar to the squeezing phase, the phase of the displacement can only be inferred by defining an external phase reference.

\subsection{Displaced squeezed single-mode Gaussian states}\label{sec:SingleDST}
For a state exhibiting nonvanishing squeezing and displacement, we can formulate different equivalent $g^{(2)}$-$g^{(3)}$ relations due to the additional dependencies on the state parameters. Thus, we need further information to check for evidence if a state satisfies a corresponding relation. One option would be to measure two out of the three parameters $|\alpha|$, $|\cov{}|$, and $\arg(\alpha^2 \cov{}^*) = 2\varphi - \theta$ with $\varphi$ being the phase of the displacement parameter. This might be done via homodyne detection which, however, can be experimentally challenging as it requires a perfectly mode-matched local oscillator and stable phase locking. Alternatively, one could engineer a nonlinear process (e.g., up-conversion, down-conversion, two-photon absorption) that converts $\hat{a}^2$ into a measurable photocurrent. While this directly accesses $\cov{}$, it demands a controllable pump laser, nonlinear crystals, and careful phase matching. Besides these methods which require additional sophisticated experimental resources, we can determine these parameters also by methods relying on the same tools used for intensity correlations, i.e., passive linear-optics networks and intensity measurements or click detectors. 
For instance, one could determine higher-order correlation functions via similar decomposition strategies as for $g^{(2)}$ and $g^{(3)}$. We would then need intensity correlations up to the $5$-th order for an unambiguous relation by using three of the four correlation functions $g^{(2)}$, $g^{(3)}$, $g^{(4)}$, and $g^{(5)}$ to determine the three unknowns $|\alpha|^2/\bar{n}$, $|\cov{}|/\bar{n}$, and $\cos(2\varphi - \theta)$. The remaining correlation function can then be expressed in terms of the other three allowing for a nontrivial test for a general single-mode Gaussian state.

Alternatively, one could interfere the state with an identically prepared state on a balanced beam splitter yielding two uncorrelated Gaussian output states. This eliminates the displacement in one output while the covariances (squeezing and thermal mean) remain unchanged. The second output is structurally identical to the original state, but its displacement amplitude is increased by a factor of $\sqrt{2}$. We can then test for a non-displaced Gaussian state via Eq.~\eqref{eq:g3-g2singleSV} in the output where the displacement is eliminated and determine $r$ and $N$ as discussed above. As a next step, we are able to determine $|\alpha|$ by measuring the mean photon number either on the original state or using the second output port of the beam splitter. If $r$, $N$, and $|\alpha|$ are known, we can determine the cosine of the relative phase between displacement and squeezing via $g^{(2)}$ measurements of the original Gaussian state or using again the state of the second output (with rescaled $\alpha$), c.f. Eq.~\eqref{eq:g2SingleMode}. Performing a $g^{(3)}$ measurement in addition could serve as a nontrivial test to overconstrain the inferred parameter set and validate the Gaussian-state model. Apart from measuring the mean photon number, we can also infer $|\alpha|$ and $\cos(2\varphi - \theta)$ from measuring $g^{(2)}$ and the no-click probability, 
\begin{align}
 p_0 &= \frac{ \E^{- \frac{1+(2N+1)\cosh(2r) + (2N+1)\sinh(2r)\cos(2\varphi-\theta)}{2N^2+2(2N+1)\cosh^2 r} |\alpha|^2} }{\sqrt{N^2+(2N+1)\cosh^2 r}}.
\label{eq:noclickSingleDST}
\end{align}
As $p_0$ also contains the relative phase, we might also combine this measurement with measuring the mean photon number. 

Thus, in summary, we can fully fix any pure or mixed single-mode Gaussian state up to a $\mathbb{Z}_2$ symmetry that mirrors the noise ellipse through the displacement axis only by intensity measurements or click detectors. This last binary freedom (resulting from the ambiguity of the cosine function) cannot be fixed by any photon-statistics measurement and needs a phase reference to be constrained for single-mode states.

As a further alternative, one can perform the hypothesis test and extract all relevant state parameters already via $g^{(2)}$ and $g^{(3)}$ measurements as well as measuring $\bar{n}$ if it is possible to manipulate one of the (source) parameters but keeping the others fixed. For instance, consider a displaced squeezed vacuum state being generated via a seeded parametric down-conversion process. Keeping the intensity of the pump laser and the intensity of the seeding laser constant but altering the relative phase between the two lasers will result in displaced squeezed states with the same $|\alpha|$ and $r$ but different relative phases between squeezing and displacement parameters. Eliminating $\mathrm{Re}\big( \alpha^2 \cov{}^* \big)$ in Eq.~\eqref{eq:g3SingleMode} by $g^{(2)}$ gives Eq.~\eqref{eq:g3-g2singleDST} which can be viewed as a linear relation, 
\begin{align}
 g^{(3)} = m\bigg(\frac{|\alpha|^2}{\bar{n}} \bigg) g^{(2)} + c\bigg(\frac{|\alpha|^2}{\bar{n}}, \frac{|\cov{}|}{\bar{n}} \bigg),
\end{align}
where the slope $m$ and intercept $c$ depend parametrically on $\frac{|\alpha|^2}{\bar{n}}$ and $\frac{|\cov{}|}{\bar{n}}$. Scanning through different $g^{(2)}$ and $g^{(3)}$ values while maintaining $m$ and $c$ constant via altering the relative phase $2\varphi - \theta$, we can measure $m$ and $c$ and thus reconstruct $|\alpha|$ and $|\cov{}|$ with the aid of the additional measurement of the mean photon number $\bar{n}$,
\begin{align}
 |\alpha|^2 &= \frac{m-9}{6}\bar{n},\\
 |\cov{}|^2 &= \bigg[\frac{(m-9)^2}{108} + \frac{c+12}{m-9}-2\bigg]\bar{n}.
\end{align}
Further, we have $|\cov{}| = (2N+1)\cosh r\, \sinh r$ and ${\bar{n} = |\alpha|^2 + \sinh^2r + N(\cosh^2r+\sinh^2r)}$, such that 
\begin{align}
 \sinh^2 r &= \frac{1}{2} \frac{2\bar{n}-2|\alpha|^{2}+1}{\sqrt{(2\bar{n}-2|\alpha|^{2}+1)^2 - 4|\cov{}|^2}} - \frac{1}{2},
 \\
 N &= \frac{1}{2} \sqrt{(2\bar{n}-2|\alpha|^{2}+1)^2 - 4|\cov{}|^2} - \frac{1}{2}.
\end{align}
Finally, we can also obtain the last remaining physical parameter describing the single-mode Gaussian state, i.e., the relative phase $2\varphi - \theta$, up to a $\mathbb{Z}_2$ reflection symmetry. Measuring a specific $g^{(2)}$ (or $g^{(3)}$) value, and knowing the ratios $|\alpha|^2/\bar{n}$ and $|\cov{}|/\bar{n}$ allows us to determine the cosine of the relative phase via Eq.~\eqref{eq:g2SingleMode} (or Eq.~\eqref{eq:g3SingleMode}), yielding
\begin{align}
 \cos(2\varphi - \theta) = \frac{\bar{n}^2 (g^{(2)} - 2)}{2|\cov{}|\, |\alpha|^2} + \frac{|\alpha|^2}{|\cov{}|} - \frac{|\cov{}|}{|\alpha|^2}.
\end{align}
As we assume that we are able to manipulate the relative phase within this scheme, one can even resolve the $\mathbb{Z}_2$ ambiguity. Performing the analysis at a different phase value allows to fix the cosine ambiguity such that we are able to also reconstruct $2\varphi-\theta$ uniquely. 
Of course, a similar analysis could also be done if another quantity of the state is experimentally accessible, i.e., $|\alpha|$ or $|\cov{}|$. Within the aforementioned example of seeded parametric down-conversion, this would correspond to altering the power of the seeding laser or the pump laser. Further, one may replace the mean photon number measurement by a click detector to determine $p_0$ and use Eq.~\eqref{eq:noclickSingleDST} in combination with the results for $g^{(2)}$ or $g^{(3)}$ to determine the state parameters on a quantitative level.

\section{Dependence of correlations and state reconstruction for Multimode Gaussian States}\label{sec:Multimode}
The analysis of potential $g^{(2)}$-$g^{(3)}$ relations and the Gaussian state reconstruction becomes more sophisticated in a multimode scenario. However, the additional degrees of freedom also offer new options for state characterization. In the single-mode case, the correlation functions $g^{(2)}$ and $g^{(3)}$ depend on precisely the same parameter combinations ($|\alpha|^{2}/\bar{n}$, $|\cov{}|/\bar{n}$, $\cos(2\varphi - \theta)$). For multimode Gaussian states, however, $g^{(3)}$ depends on more algebraic combinations involving relative phase terms among different modes that are not present for $g^{(2)}$ correlations. 
In addition, both correlation functions depend on a further second-order moment identifiable as the first-order coherence $g^{(1)}_{ij} \sim \mo{\us i j}$, which quantifies field correlations and being nontrivial only for distinct modes.

From Eq.~\eqref{eq:g2Decomposition}, it follows that any of the quantities $|g^{(1)}_{ij}|$, $\bar{n}_i$, $|\cov{ij}|$, $|\alpha_i|$, or $\mathrm{Re}(\alpha_i \alpha_j \cov{ij}^{*})$ can be expressed in terms of $g^{(2)}_{ij}$ and the remaining quantities. As in the single-mode case, a convenient choice would be a quantity that is sophisticated to measure or easy to manipulate while keeping all others fixed. Note that $\bar{n}_i$ appears only in terms of ratios with either the covariance matrix $\cov{ij}/\sqrt{\bar{n}_i \bar{n}_j}$ or the displacement $\alpha_{i}/\sqrt{\bar{n}_i}$ in both, Eq.~\ref{eq:g2Decomposition} and Eq.~\ref{eq:g3Decomposition}. Further, all $\bar{n}_i$ can be extracted from additional intensity measurements which might be required for extracting specific state information as discussed in the single-mode case. Thus, the mean photon number $\bar{n}_i$ of the different modes $i$ would not be an appropriate candidate for elimination in the following analysis.

From an operational perspective, measurements of $g^{(1)}_{ij}$ are also comparatively straightforward, as they can be realized with the same passive linear-optical resources already required for intensity-correlation experiments. The only additional element is an interferometric stage that imposes a calibrated phase shift between modes $i$ and $j$, e.g., via an electro-optic modulator. The output intensities yield the fringe visibility $\sim |g^{(1)}_{ij}|$, e.g., for a balanced Mach-Zehnder interferometer. Repeating the measurement with known phase shifts in one of the arms, we can extract the phase of $g^{(1)}_{ij}$ being the fringe phase. Amplitude and phase of $g^{(1)}_{ij}$ can thus be extracted without introducing a local oscillator or invoking any nonlinear process. 
Therefore, we only consider to eliminate one of the ratios $|\cov{ij}|/\sqrt{\bar{n}_i \bar{n}_j}$ or $|\alpha_i|/\sqrt{\bar{n}_i}$, or the relative phases between squeezing and displacement encoded in $\mathrm{Re}(\alpha_i \alpha_j \cov{ij}^{*})/(\bar{n}_i \bar{n}_j)$ via $g^{(2)}$ measurements. Thus, we are able to organize the discussion in the same manner as in the single-mode case, focusing on states without squeezing ($\cov{ij}=0$), states without displacement ($\alpha_i=0$), and states with nonvanishing displacement and squeezing.

For the following analysis, we assume access to measurements of the intensity correlations $g^{(2)}_{ij}$ and $g^{(3)}_{ijk}$, the first-order field correlations $g^{(1)}_{ij}$, and the mean photon numbers $\bar{n}_i$ of each mode. Instead of using the mean photon numbers as loss-sensitive observables, one may equivalently use the (no-)click probabilities of each mode, $p_{0,i} = \tr \big( \hat\rho_\mathrm{g} \op{0}_i \big)$, which also encode all relevant information that is needed for state reconstruction. However, for simplicity, we will only focus on measuring mean photon numbers instead of click statistics in what follows.

\subsection{Multimode Displaced Thermal States}
First, we consider multimode Gaussian states without squeezing, implying $\cov{ij}=0$. The most general state in this class is a displaced thermal state being parametrized by $\hat{D}(\vec\alpha)\hat{R}(\boldsymbol{\phi}) \,\hat\rho_\mathrm{th,M}\, \hat{R}^\dagger(\boldsymbol{\phi})\hat{D}^\dagger(\vec\alpha)$, including multimode displaced vacuum states ($N_i = 0$ for all modes) as well as thermal states ($\vec\alpha = 0$).

The second-order correlation functions for this class of states read,
\begin{align}
 g^{(2)}_{ij} &= 1 + |g_{ij}^{(1)}|^2-\frac{|\alpha_i|^2|\alpha_j|^2}{\bar{n}_i\bar{n}_j}.
\end{align}
Considering the diagonal elements $g^{(2)}_{ii}$, it is possible to eliminate the moduli of the displacements, 
\begin{align}
 |\alpha_i|^2 = \bar{n}_i  \sqrt{2-g_{ii}^{(2)}}.
\label{eq:alphaMultiDT}
\end{align}
In contrast to the single-mode case, the multimode scenario allows for new options regarding hypothesis testing. If $|g_{ij}^{(1)}|$ is known, we can perform a nontrivial test already at the level of the second-order correlation functions with the aid of the off-diagonal elements $i \neq j$ as 
\begin{align}
g^{(2)}_{ij} = 1 + |g^{(1)}_{ij}|^2 - \sqrt{\big(2-g_{ii}^{(2)}\big)\big(2-g_{jj}^{(2)}\big)}.
\label{eq:g2g1RelationDTh}
\end{align} 
Note that this is a straightforward generalization of the well-known relation ${g^{(2)}_{ij} = 1 + |g_{ij}^{(1)}|^2}$ for thermal states without any displacement or squeezing. 

To further strengthen the evidence for non-squeezed multimode Gaussian states, we could incorporate $g^{(3)}$ measurements into the analysis. 
Also the third-order correlations simplify substantially if $\cov{ij}=0$ in Eq.~\eqref{eq:g3Decomposition}:
\begin{align}
        g^{(3)}_{ijk} &= 1 + |g_{ij}^{(1)}|^2 + |g_{jk}^{(1)}|^2 +|g_{ik}^{(1)}|^2 + 2\mathrm{Re}[g^{(1)}_{ij}g_{jk}^{(1)}g_{ki}^{(1)}]
        \notag \\
        &\quad -\frac{|\alpha_i\alpha_j|^2}{\bar{n}_i\bar{n}_j} -\frac{|\alpha_j\alpha_k|^2}{\bar{n}_j\bar{n}_k} -\frac{|\alpha_i\alpha_k|^2}{\bar{n}_i\bar{n}_k }
        + 4\frac{|\alpha_i\alpha_j\alpha_k|^2}{\bar{n}_i\bar{n}_j\bar{n}_k} 
        \notag \\
        &\quad- 2\frac{|\alpha_i|^2}{\bar{n}_i}\frac{\mathrm{Re}[g_{jk}^{(1)}\alpha_j\alpha_k^*]}{\sqrt{\bar{n}_j\bar{n}_k}}
         - 2\frac{|\alpha_j|^2}{\bar{n}_j}\frac{\mathrm{Re}[g_{ki}^{(1)}\alpha_k\alpha_i^*]}{\sqrt{\bar{n}_i\bar{n}_k}}
        \notag \\
        &\quad
        - 2\frac{|\alpha_k|^2}{\bar{n}_k}\frac{\mathrm{Re}[g_{ij}^{(1)}\alpha_i\alpha_j^*]}{\sqrt{\bar{n}_i\bar{n}_j}}.
        \label{eq:g3multiDT}
\end{align}

Plugging Eq.~\eqref{eq:alphaMultiDT} into Eq.~\eqref{eq:g3multiDT}, we have a fixed relation of $g^{(3)}_{ijk}$ depending on $g^{(2)}$ and $g^{(1)}$ measurements, 
\begin{widetext}
\begin{align}
  g^{(3)}_{ijk} &= g_{ij}^{(2)} + g_{jk}^{(2)} + g_{ik}^{(2)} - 2  
  + 2|g_{ij}^{(1)}| \, |g_{jk}^{(1)}| \, |g_{ki}^{(1)}|\cos(\Phi_{ij}+\Phi_{jk}-\Phi_{ik})
  + 4\sqrt{\big(2-g_{ii}^{(2)}\big)\big(2-g_{jj}^{(2)}\big)\big(2-g_{kk}^{(2)}\big)} 
  \notag \\
  &\quad - 2\sqrt[4]{\big(2-g_{ii}^{(2)}\big)\big(2-g_{jj}^{(2)}\big)\big(2-g_{kk}^{(2)}\big)} 
  \bigg[ |g_{ij}^{(1)}| \sqrt[4]{2-g_{kk}^{(2)}} \cos(\Phi_{ij}+\varphi_i -\varphi_j) 
  + |g_{jk}^{(1)}|\sqrt[4]{2-g_{ii}^{(2)}}\cos(\Phi_{jk}+\varphi_j -\varphi_k) 
  \notag \\
  &\qquad\qquad\qquad\qquad\qquad\qquad\qquad\qquad\quad + |g_{ik}^{(1)}|\sqrt[4]{2-g_{jj}^{(2)}}\cos(\Phi_{ik}+\varphi_i -\varphi_k)
  \bigg],
  \label{eq:g3-g2multiDT}
\end{align}
\end{widetext}
where $\varphi_i$ denote the phases of the displacements ${\alpha_i = |\alpha_i| \E^{\im \varphi_i}}$ and $\Phi_{ij}$ are the phases of the field correlations, $g^{(1)}_{ij} = |g_{ij}^{(1)}|\E^{\im \Phi_{ij}}$ with $\Phi_{ij} = -\Phi_{ji}$.
The only unknown quantities on the right-hand side of Eq.~\eqref{eq:g3multiDT} are relative phases of the displacements. If we are able to scan this relation through different parameters, e.g., $|\alpha_i|$, but keeping these relative phases fixed, we have an additional tool to obtain evidence for (or against) a multimode displaced thermal state. However, due to the additional multimode structures, we can also perform nontrivial hypothesis tests without sophisticated parameter scans. As a first consistency check, we can test the $g^{(2)}$-$g^{(3)}$ relation for $i=j=k$ for the $M$ modes individually where the phase dependency drops out and Eq.~\eqref{eq:g3-g2multiDT} reduces to Eq.~\eqref{eq:g3-g2singleDT}. Further, we can remove the phase dependencies in Eq.~\eqref{eq:g3-g2multiDT} by using a subset of $g^{(3)}$ measurements. For instance, we can infer $\cos(\Phi_{ij}+\varphi_i-\varphi_j)$ by considering the third-order correlation functions with two matching indices,
\begin{align}
 g_{iij}^{(3)} &= 2g_{ij}^{(2)} + 2|g_{ij}^{(1)}|^2 - 1 \notag 
 \\
 &\quad - \sqrt{\big(2-g_{ii}^{(2)}\big)\big(2-g_{jj}^{(2)}\big)}\Big(2 - \sqrt{2-g_{ii}^{(2)}}\Big) 
 \label{eq:giijDT} \\
 &\quad -4|g_{ij}^{(1)}|\sqrt[4]{\big(2-g^{(2)}_{ii}\big)^3\big(2-g_{jj}^{(2)}\big)}\cos(\Phi_{ij}+\varphi_i-\varphi_j). \notag
\end{align}
Note that $g_{jji}^{(3)}$ depends on precisely the same cosine term due to the antisymmetry of $\Phi_{ij}$ and the property that the cosine is an even function, $\cos(\Phi_{ji}+\varphi_j-\varphi_i) = \cos(\Phi_{ij}+\varphi_i-\varphi_j)$. Thus comparing $g_{iij}^{(3)}$ and $g_{jji}^{(3)}$ measurements provides a nontrivial test for the class of multimode displaced thermal states. Further, we can insert the relation for the cosine of the relative phase combinations derived from Eq.~\eqref{eq:giijDT} into Eq.~\eqref{eq:g3-g2multiDT}, relating $g^{(3)}$ with $g^{(2)}$ and $|g^{(1)}|$. Finally, we could also remove the $|g^{(1)}|$ dependency via Eq.~\eqref{eq:g2g1RelationDTh} such that we obtain a relation combining only $g^{(3)}$ and $g^{(2)}$ measurements. Thus, we are able to obtain strong evidence for this particular class of states even without measuring $g^{(1)}_{ij}$.

Furthermore, we are able to perform a nontrivial hypothesis test by actually extracting the phases of the displacement parameters from correlation measurements. 
In principle, measurements of $g^{(1)}_{ij}, g^{(2)}_{ij}$, and $g^{(3)}_{ijk}$, provide access to all relative phases $\varphi_i - \varphi_j$, which can be inferred from Eq.~\eqref{eq:giijDT}.
If these correlation measurements permit a self-consistent solution, we obtain evidence that the state belongs to the class of multimode displaced thermal states. In this scenario, the actual phases $\varphi_i$ can be determined uniquely up to a specific ambiguity. First note that Eq.~\eqref{eq:giijDT} is invariant under a global phase shift $\psi_{\alpha}$ of the displacement, i.e., $\varphi_i \to \varphi_i + \psi_{\alpha}$ for all modes $i$. Due to this $\mathrm{U}(1)$ symmetry, we are free to choose $\varphi_1 = 0$ without loss of generality, which corresponds to our freedom of choosing the overall phase reference, i.e., the origin of the angular coordinates in phase space. Then, we can determine the phases $\varphi_i$ from $g_{11i}^{(3)}$ up to the cosine ambiguity for each mode $i\in\{2,\cdots,M\}$. However, note that these binary ambiguities are strongly constrained, since the relation given in Eq.~\eqref{eq:giijDT} must also be satisfied for all $i\neq 1$ and $i \neq j \neq 1$.

One can show that in most cases, the resulting system of equations, $c_{ij} = \cos(\Phi_{ij}+\varphi_i-\varphi_j)$, has either zero or one solution depending on the phases $\Phi_{ij}$ and the real numbers $c_{ij}$ that can be extracted from Eq.~\eqref{eq:giijDT}. Additional solutions may only arise in highly symmetric edge cases, see App.~\ref{app:phases} for further details. For example, in case of all $\Phi_{ij} = 0$, we have the additional freedom to flip signs of all phases simultaneously, $\varphi_i \to -\varphi_i$, leaving the cosine terms in Eq.~\eqref{eq:giijDT} invariant. However, for almost all practical cases containing noisy measurement data, these nongeneric degeneracies are lifted as any perturbation of the edge case data will lead to a unique solution (up to the global phase shift if we do not fix the reference system). Thus, if no consistent solution exists that satisfies all constraints, the state cannot be a multimode Gaussian state without squeezing. If the state belongs to this class, we will usually find one solution. Only, the two mode case exhibits an additional $\mathbb{Z}_{2}$ ambiguity in general. In this scenario, we are simply lacking the additional relations of the higher dimensional structures to further constrain $\varphi_1 - \varphi_2 = -\Phi_{12} \pm \arccos c_{12}$.

Apart from verifying evidence for this class of states, we are also able to almost fully reconstruct the entire state similar to the single-mode case. Besides measuring correlation functions ($g^{(1)}$, $g^{(2)}$, and $g^{(3)}$), the only additional measurements needed for this purpose are the mean photon numbers in the individual modes $\bar{n}_i$. For a general displaced thermal state, the mean photon numbers and first-order correlation functions read:
\begin{align}
        \bar{n}_i &= \big(\E^{\im\boldsymbol \phi}\boldsymbol{\mathcal{D}}\E^{-\im\boldsymbol \phi}\big)_{ii} + |\alpha_i|^2,
        \label{eq:meanPhotonDT}\\
        g_{ij}^{(1)} &= \frac{\big(\E^{\im\boldsymbol \phi}\boldsymbol{\mathcal{D}}\E^{-\im\boldsymbol \phi}\big)_{ji} + \alpha_i^*\alpha_j}{\sqrt{\bar{n}_i\bar{n}_j}},
        \label{eq:g1DT}
\end{align}
with $\boldsymbol{\mathcal{D}} = \mathrm{diag}[N_1,\cdots, N_M]$ and $\boldsymbol{\phi}$ encoding the parameters describing the mixing of the different modes via $\hat{R}(\boldsymbol{\phi})$. 
As already discussed, the displacement magnitudes $|\alpha_i|$ can be extracted from $g^{(2)}_{ii}$ measurements if $\bar{n}_i$ is known, see Eq.~\eqref{eq:alphaMultiDT}. Further, we are able to determine the phases of the displacements up to a global phase shift (and potential $\mathbb{Z}_2$ ambiguities for special cases). 
Finally, we can reconstruct the thermal excitations and mode rotations. By measuring $\bar{n}_{i}$ and $g^{(1)}_{ij}$ and using the reconstructed displacements $\alpha_i$, we can compute the diagonal and off-diagonal elements of $\big(\E^{\im\boldsymbol \phi}\boldsymbol{\mathcal{D}}\E^{-\im\boldsymbol \phi}\big)$ via Eq.~\eqref{eq:meanPhotonDT} and Eq.~\eqref{eq:g1DT}, respectively, which can be written in compact form as $\big(\E^{\im\boldsymbol{\phi}}\boldsymbol{\mathcal{D}}\E^{-\im\boldsymbol{\phi}}\big)_{ji} = \sqrt{\bar{n}_i\bar{n}_j}g_{ij}^{(1)} - \alpha_i^*\alpha_j$. Note that the phase ambiguity of the displacements precisely cancel as the matrix on the right-hand side only depends on relative phases $\varphi_i - \varphi_j$. This matrix is clearly diagonalizable with eigenvalues $N_i$ and eigenvectors constituting the matrix $\E^{\im\boldsymbol{\phi}}$. 
This spectral decomposition is unique up to rotations between modes with same eigenvalues. In case of no degeneracies in $\boldsymbol{\mathcal{D}}$, i.e., non-repeating $N_i$, the decomposition is unique up to local phase rotations $\boldsymbol{Q} = \mathrm{diag}[\E^{\im\psi_1},\cdots,\E^{\im\psi_M}]$ being generated by elements of the Cartan subalgebra of $\mathrm{u}(M)$. This freedom reflects the degree of freedom of local mode phases within the rotation operator $\hat{R}(\boldsymbol{\phi})$ which do not alter the density matrix of the thermal state given in Eq.~\eqref{eq:thermalState} and are therefore physically irrelevant. Hence, we are able to reconstruct the displacement $\vec{\alpha}$ (up to a global phase), the thermal excitations encoded in the diagonal matrix $\boldsymbol{\mathcal{D}}$ and the mode rotations apart from physically irrelevant local rotations. The global displacement phase could also be expressed as a rather trivial additional rotation operation of the form $\hat{R}(\E^{\im\psi_\alpha}\mathds{1})$.

As a final remark, we would like to emphasize that these results also contain well-known limiting cases as special examples.
This includes the class of multimode displaced vacuum states if all $N_i = 0$. However, this special case is trivial as the modes of such a state are uncorrelated. Further, $\bar{n}_i = |\alpha_i|^2$ such that $g^{(2)}_{ij} = 1$ and $g^{(3)}_{ijk} = 1$. The displacement magnitudes can be obtained from measuring the mean photon number of each mode and the relative phases of the displacement parameters can be inferred from measuring the phases of the field correlator $g^{(1)}_{ij} = \E^{\im (\varphi_j-\varphi_i)}$. This fully fixes the state up to an absolute phase corresponding to a global rotation in phase space. Further, we obtain the known relations $g^{(2)}_{ij} = 1 + |g^{(1)}_{ij}|^2$ and $g^{(3)}_{ijk} = 1 + |g^{(1)}_{ij}|^2 +  |g_{jk}^{(1)}|^2 +|g_{ik}^{(1)}|^2 + 2\mathrm{Re}[g^{(1)}_{ij}g_{jk}^{(1)}g_{ki}^{(1)}]$ for thermal states. Finally, we obtain the result of the single-mode case for each mode if $\boldsymbol{\phi} = 0$ as all modes become uncorrelated in this limit.

\subsection{Multimode Squeezed Thermal States}\label{sec:MultiST}
For the class of multimode Gaussian states without displacement, the most general state is represented as $\hat{S}(\boldsymbol{z})\hat{R}(\boldsymbol{\phi}) \,\hat\rho_\mathrm{th,M}\, \hat{R}^\dagger(\boldsymbol{\phi})\hat{S}^\dagger(\boldsymbol{z})$ being an arbitrary squeezed thermal state.
The second-order correlation function for such a state is given by 
\begin{align}
 g^{(2)}_{ij} &= 1 + |g^{(1)}_{ij}|^2 + \frac{|\cov{ij}|^2}{\bar{n}_i\bar{n}_j}.
\end{align}
In contrast to the non-squeezed scenario, we are not able to perform a nontrivial hypothesis test at the level of $g^{(2)}$ and $g^{(1)}$ measurements as we need all $\frac{1}{2}M(M+1)$ independent measurements $g^{(2)}_{ij}$ and knowing  the $\frac{1}{2}M(M-1)$ elements of $|g^{(1)}_{ij}|$ to fix the $\frac{1}{2}M(M+1)$ ratios $|\cov{ij}|^2/(\bar{n}_i\bar{n}_j)$.

For a general squeezed thermal state, the third-order correlations simplify to
\begin{widetext}
\begin{align}
 g^{(3)}_{ijk} &= 1 + |g_{ij}^{(1)}|^2 + |g_{jk}^{(1)}|^2 +|g_{ik}^{(1)}|^2 + 2\mathrm{Re}\big[g^{(1)}_{ij}g_{jk}^{(1)}g_{ki}^{(1)}\big] 
 + \frac{|\cov{ij}|^2}{\bar{n}_i\bar{n}_j} + \frac{|\cov{jk}|^2}{\bar{n}_j\bar{n}_k} + \frac{|\cov{ik}|^2}{\bar{n}_i\bar{n}_k} \notag\\
 &\quad  + 2\frac{\mathrm{Re}[g_{ij}^{(1)}\cov{jk}^*\cov{ik}]}{\bar{n}_k\sqrt{\bar{n}_i\bar{n}_j}}
               + 2\frac{\mathrm{Re}[g_{jk}^{(1)}\cov{ik}^*\cov{ij}]}{\bar{n}_i\sqrt{\bar{n}_j\bar{n}_k}}
 	      + 2\frac{\mathrm{Re}[g_{ki}^{(1)}\cov{ij}^*\cov{jk}]}{\bar{n}_j\sqrt{\bar{n}_i\bar{n}_k}}.
\label{eq:g3-g2multiST}
\end{align}
If both $g^{(2)}$ and $g^{(1)}$ are experimentally accessible, we can determine $g^{(3)}$ as a function of $g^{(2)}$, $g^{(1)}$, and phase terms by substituting $|\cov{ij}|^2/(\bar{n}_i\bar{n}_j) = g^{(2)}_{ij}-|g^{(1)}_{ij}|^2-1$: 
    \begin{align}
        g^{(3)}_{ijk} &= g_{ij}^{(2)} + g_{jk}^{(2)} + g_{ik}^{(2)} - 2 + 2|g^{(1)}_{ij}||g_{jk}^{(1)}||g_{ik}^{(1)}|\cos(\Phi_{ij}+\Phi_{jk}-\Phi_{ik}) \notag\\
        &\quad + 2 |g_{ij}^{(1)}|\sqrt{\big(g^{(2)}_{jk}-|g_{jk}^{(1)}|^2-1 \big)\big(g^{(2)}_{ik}-|g_{ik}^{(1)}|^2-1\big)}\cos(\Phi_{ij}-\Theta_{jk}+\Theta_{ik}) \notag\\
        &\quad + 2 |g_{jk}^{(1)}|\sqrt{\big(g^{(2)}_{ik}-|g_{ik}^{(1)}|^2-1\big)\big(g^{(2)}_{ij}-|g_{ij}^{(1)}|^2-1\big)}\cos(\Phi_{jk}-\Theta_{ik}+\Theta_{ij}) \notag\\
        &\quad + 2 |g_{ik }^{(1)}|\sqrt{\big(g^{(2)}_{ij}-|g_{ij}^{(1)}|^2-1\big)\big(g^{(2)}_{jk}-|g_{jk}^{(1)}|^2-1\big)}\cos(\Phi_{ki}-\Theta_{ij}+\Theta_{jk}),
\label{eq:g2g3MultiST}
\end{align}
\end{widetext}
where $\Theta_{mn}$ are the phases of the covariance elements ${\cov{mn}=|\cov{mn}|\E^{\im\Theta_{mn}}}$. To perform a nontrivial hypothesis test, we can test whether the resulting system of equations has self-consistent solutions for the covariance phases $\Theta_{mn}$. Similar to the case of extracting the phases for displacements in the previous section, the system of equations for $\Theta_{mn}$ is overconstrained for $M>2$. 
For this analysis, it is again convenient to investigate third-order correlations for two matching indices,
\begin{align}
        g_{iij}^{(3)} &=  g_{ii}^{(2)} + 4g_{ij}^{(2)} - 4 \notag\\
        &\quad + 4|g_{ij}^{(1)}|\sqrt{(g^{(2)}_{ij}-|g_{ij}^{(1)}|^2-1)(g^{(2)}_{ii}-2)}  \notag\\
        &\qquad\times\cos(\Phi_{ij}-\Theta_{ij}+\Theta_{ii}),
\end{align}
from which we can extract the relative phases $\Theta_{ii} - \Theta_{ij}$. As the covariance matrix is symmetric, we can determine all $\frac{1}{2}M(M+1)$ phases $\Theta_{ij}$ from an appropriately chosen subset of $g_{iij}^{(3)}$ measurements apart from a global $\mathrm{U}(1)$ symmetry $(\Theta_{ij} \to \Theta_{ij} + \psi_{\mathrm{z}})$ and potential $\mathbb{Z}_2$ ambiguities for special cases, see App.~\ref{app:phases}. If a consistent solution exists, one obtains strong evidence for a multimode Gaussian state without displacement. This evidence can be further substantiated by checking the nontrivial relations of the correlation functions for further index combinations, e.g., $g^{(3)}_{iii} = 9 g^{(2)}_{ii} - 12$ or the general form for three nonmatching indices given by Eq.~\eqref{eq:g2g3MultiST}. 

There is only one peculiarity in the case $M=2$. For two-mode systems, one generically obtains four discrete solutions apart from the additional continuous global phase shift. The reason for these four solutions is simply given by the fact that only two nontrivial $g^{(3)}$ measurements contain relative phases ($g^{(3)}_{112}$ and $g^{(3)}_{221}$) such that we are not able to constrain the freedom of the two sign choices associated with the corresponding cosine functions, resulting in four independent solutions. 

Nevertheless, for most practical cases, we can uniquely reconstruct all phases of the covariance matrix for $M\geq 3$ up to a global rotation that translates into a global $\mathrm{U}(1)$ freedom of the squeezing phases $\boldsymbol{\theta}$ reflecting the freedom to choose the reference system. Moreover, we can also fully determine the moduli of the covariance elements via correlation measurements and measuring the mean photon numbers,
\begin{align}
 |\cov{ij}|= \sqrt{\bar{n}_i\bar{n}_j \big(g^{(2)}_{ij}-|g^{(1)}_{ij}|^2-1\big)},
\end{align}
such that we can fully reconstruct $\cov{ij}$. 

Knowing the covariance matrix, all $g^{(1)}_{ij}$, and the mean photon numbers in each mode $\bar{n}_{i}$, enables a reconstruction of the parameters characterizing a general multimode squeezed thermal state, i.e., the squeezing matrix $\boldsymbol{z}$, the rotation $\boldsymbol{\phi}$, as well as the thermal excitations $N_i$. For vanishing displacements, the mean photon numbers, covariances and first-order correlation functions of a Gaussian state read:
\begin{align}
        \bar{n}_i &= \sinh^2(\boldsymbol{r})_{ii} + \big(\cosh(\boldsymbol{r})\E^{\im\boldsymbol{\phi}}\boldsymbol{\mathcal{D}}\E^{-\im\boldsymbol{\phi}}\cosh(\boldsymbol{r})\big)_{ii} \notag\\
        &\quad+ \big(\sinh(\boldsymbol{r})\E^{\im\boldsymbol{\theta}}\E^{-\im\boldsymbol{\phi}^\top}\boldsymbol{\mathcal{D}}\E^{\im\boldsymbol{\phi}^\top}\E^{-\im\boldsymbol{\theta}}\sinh(\boldsymbol{r})\big)_{ii}, 
\label{eq:MultiSTmeannumber}
        \\
        \cov{ij} &= \frac{1}{2}\big(\sinh(2\boldsymbol r)\E^{\im\boldsymbol \theta}\big)_{ij} \notag\\
        &\quad + \big(\sinh(\boldsymbol r)\E^{\im\boldsymbol \theta} \E^{-\im \boldsymbol{\phi}^\top}\boldsymbol{\mathcal{D}}\E^{\im\boldsymbol{\phi}^\top}\cosh(\boldsymbol r^\top)\big)_{ij} \notag\\
        &\quad + \big(\cosh(\boldsymbol r) \E^{\im\boldsymbol\phi}\boldsymbol{\mathcal{D}}\E^{-\im\boldsymbol\phi}\E^{\im\boldsymbol\theta^\top}\sinh(\boldsymbol{r}^\top)\big)_{ij},
        \\
        g^{(1)}_{ij} &= \frac{\sinh^2(\boldsymbol{r})_{ji} + \big(\cosh(\boldsymbol{r})\E^{\im\boldsymbol{\phi}}\boldsymbol{\mathcal{D}}\E^{-\im\boldsymbol{\phi}}\cosh(\boldsymbol{r})\big)_{ji}}{\sqrt{\bar{n}_i\bar{n}_j}} \notag\\
        &\quad + \frac{\big(\sinh(\boldsymbol{r})\E^{\im\boldsymbol{\theta}}\E^{-\im\boldsymbol{\phi}^\top}\boldsymbol{\mathcal{D}}\E^{\im\boldsymbol{\phi}^\top}\E^{-\im\boldsymbol{\theta}}\sinh(\boldsymbol{r})\big)_{ji}}{\sqrt{\bar{n}_i\bar{n}_j}}.
\end{align}
From these relations, we can reconstruct the full covariance matrix of annihilation and creation operators
\begin{align}
        \boldsymbol{V}^{(c)} &= 
        \begin{pmatrix}
        \boldsymbol{A}  & \boldsymbol{B}\\
        \boldsymbol{B}^*& \boldsymbol{A}^*
        \end{pmatrix},
        &
        \begin{aligned}
        A_{ij} &= \tfrac{1}{2}\langle\{\hat{a}_i,\hat{a}_j^\dagger\}\rangle - \langle\hat{a}_i\rangle\langle\hat{a}_j^\dagger\rangle \\
        B_{ij} &= \langle\hat{a}_i\hat{a}_j\rangle-\langle\hat{a}_i\rangle\langle\hat{a}_j\rangle.
        \end{aligned}
\end{align}
This matrix can be converted into the real covariance matrix $\boldsymbol{V}^{(r)}$ which corresponds to switching from a ladder operator picture to a momentum and position representation \cite{vallone_means_2019}. The matrix $\boldsymbol{V}^{(r)}$ is real, symmetric, and positive definite, which permits the application of the Williamson decomposition to extract the thermal excitations, rotations, and squeezing parameters, up to local rotations within degenerate eigenspaces \cite{simon_quantumnoise_1994,vallone_means_2019,houde_matrix_2024}. The decomposition
\begin{equation}
    \boldsymbol{V}^{(r)} = \boldsymbol{U}_r \boldsymbol{D}\,\boldsymbol{U}_r^\top
\end{equation}
yields the thermal excitations directly from the symplectic eigenvalues $D_i = N_i + \tfrac{1}{2}$, while the symplectic matrix
\begin{equation}
    \boldsymbol{U}_r =
    \begin{pmatrix}
        \mathrm{Re}[\boldsymbol{E}+\boldsymbol{F}] & -\mathrm{Im}[\boldsymbol{E}-\boldsymbol{F}] \\[4pt]
        \mathrm{Im}[\boldsymbol{E}+\boldsymbol{F}] & \mathrm{Re}[\boldsymbol{E}-\boldsymbol{F}]
    \end{pmatrix}
\end{equation}
encodes the information required to reconstruct $\boldsymbol{\phi}$ and $\boldsymbol{z}$, cf. Eq.~\eqref{eq.6}. The decomposition is unique up to a degenrate mode rotation $\boldsymbol{Q} = \mathrm{diag}[\E^{\im\psi_1},\cdots,\E^{\im\psi_M}]$,
\begin{equation}
    \boldsymbol{U}_r \;\to\; \boldsymbol{U}_r
    \begin{pmatrix}
        \mathrm{Re}[\boldsymbol{Q}] & -\mathrm{Im}[\boldsymbol{Q}] \\[4pt]
        \mathrm{Im}[\boldsymbol{Q}] & \mathrm{Re}[\boldsymbol{Q}]
    \end{pmatrix},
\end{equation}
which acts locally on non-degenerate modes or mixes degenerate ones, but does not affect any physically relevant quantities as it corresponds to global phase shifts which leave the density matrix invariant. Thus, the Williamson decomposition allows to extract all relevant state parameters and characterize the state fully up to a global rotation in phase space. 

\subsection{Multimode Displaced Squeezed Thermal States}
Considering the most general Gaussian state, being a multimode displaced squeezed thermal state (see Eq.~\eqref{eq:multiGaussian}), we have to consider the full expressions for $g^{(2)}$ and $g^{(3)}$ given by Eqs.~\eqref{eq:g2Decomposition} and \eqref{eq:g3Decomposition}, respectively. Studying an explicit $g^{(2)}$-$g^{(3)}$ relation does not necessarily yield additional insights. Moreover, there are several versions of such a relation due to the dependency of $g^{(2)}$ on multiple state parameters as discussed above. However, a general explicit relation is not needed for state classification and reconstruction as we can exploit the various degrees of freedom in a multimode scenario to over constrain the state reconstruction providing nontrivial consistency checks as in the previous two subsections. 
The required expressions for $\bar{n}_i$, $g^{(1)}_{ij}$, and $\cov{ij}$ can be calculated straightforwardly. While the covariance matrix has the same form as in the squeezed-thermal case, the mean photon numbers acquire an additional contribution $|\alpha_i|^2$, reflecting the displacement of each mode. Also, the first-order correlation functions receive an additional displacement contribution. More precisely, we obtain
\begin{widetext}
\begin{align}
        \bar{n}_i &= |\alpha_i|^2 + \sinh^2(\boldsymbol{r})_{ii} + \big(\cosh(\boldsymbol{r})\E^{\im\boldsymbol{\phi}}\boldsymbol{\mathcal{D}}\E^{-\im\boldsymbol{\phi}}\cosh(\boldsymbol{r})\big)_{ii}
       + \big(\sinh(\boldsymbol{r})\E^{\im\boldsymbol{\theta}}\E^{-\im\boldsymbol{\phi}^\top}\boldsymbol{\mathcal{D}}\E^{\im\boldsymbol{\phi}^\top}\E^{-\im\boldsymbol{\theta}}\sinh(\boldsymbol{r})\big)_{ii},
\label{eq:MultiDSTmeannumber} 
        \\
        g^{(1)}_{ij} &= \frac{\sinh^2(\boldsymbol{r})_{ji} + \big(\cosh(\boldsymbol{r})\E^{\im\boldsymbol{\phi}}\boldsymbol{\mathcal{D}}\E^{-\im\boldsymbol{\phi}}\cosh(\boldsymbol{r})\big)_{ji}}{\sqrt{\bar{n}_i\bar{n}_j}}
        + \frac{\alpha_i^*\alpha_j + \big(\sinh(\boldsymbol{r})\E^{\im\boldsymbol{\theta}}\E^{-\im\boldsymbol{\phi}^\top}\boldsymbol{\mathcal{D}}\E^{\im\boldsymbol{\phi}^\top}\E^{-\im\boldsymbol{\theta}}\sinh(\boldsymbol{r})\big)_{ji}}{\sqrt{\bar{n}_i\bar{n}_j}},
\label{eq:g1MultiDST}
        \\
        \cov{ij} &= \frac{1}{2}\big(\sinh(2 \boldsymbol r)\E^{\im\boldsymbol \theta}\big)_{ij}
        + \big(\sinh(\boldsymbol r)\E^{\im\boldsymbol \theta} \E^{-\im \boldsymbol{\phi}^\top}\boldsymbol{\mathcal{D}}\E^{\im\boldsymbol{\phi}^\top}\cosh(\boldsymbol r^\top)\big)_{ij}
        + \big(\cosh(\boldsymbol r) \E^{\im\boldsymbol\phi}\boldsymbol{\mathcal{D}}\E^{-\im\boldsymbol\phi}\E^{\im\boldsymbol\theta^\top}\sinh(\boldsymbol{r}^\top)\big)_{ij}.
\end{align}
\end{widetext}

Several strategies could be used to extract the state parameters from a subset of the correlation and mean photon number measurements. Here, we present a convenient approach based on interfering the state with an identically prepared state on a balanced beam splitter, in the same spirit as our discussion of the single-mode case, see Sec.~\ref{sec:SingleDST}. Related interferometric schemes have also been proposed to witness continuous-variable entanglement for Gaussian and non-Gaussian states \cite{callus_interferometric_2025}.
By this interference, we can fully remove the displacement in one output in a multimode setting as well and obtain two uncorrelated states with unaffected covariances. This becomes directly apparent from the perspective of the Bogoliubov transformation where the beam splitter transformation is described by 
\begin{align}
 \boldsymbol{U}_\mathrm{BS} &= 
 \frac{1}{\sqrt{2}}\begin{pmatrix} \mathds{1} & \mathds{1}\\ \mathds{1} & -\mathds{1} \end{pmatrix},
\end{align}
and the states characterized by the Bogoliubov block $\boldsymbol{L}$ (second moments) and displacement vector $\vec{A}$ (first moments) transform as \cite{weedbrook_gaussian_2012,brask_gaussian_2022}
\begin{align}
 \boldsymbol{U}_\mathrm{BS}(\boldsymbol{L}\oplus\boldsymbol{L})\boldsymbol{U}_\mathrm{BS}^\top  &= \boldsymbol{L}\oplus\boldsymbol{L}, \\
 \boldsymbol{U}_\mathrm{BS}(\vec{A}\oplus\vec{A}) &= (\sqrt{2}\vec{A})\oplus\vec{0}.
\end{align}
After this interference step, the zero-mean output state is just a multimode squeezed thermal state if the initial state was indeed a multimode displaced squeezed thermal state. Thus, evidence for such an output state can be obtained and its reconstruction carried out as discussed in Sec.~\ref{sec:MultiST}. Finally, the displacement parameters of the original state can be determined by performing measurements in the other output port of the beam splitter or on the original state emitted by the source. To determine the moduli $|\alpha_i|$, we compare the mean photon number $\bar{n}_i$ of the original state (or the other output state with displacements $\sqrt{2}|\alpha_i|$) to the measurements of $\bar{n}_i$ in the zero-mean output, cf. Eqs.~\eqref{eq:MultiDSTmeannumber} and \eqref{eq:MultiSTmeannumber}. The only remaining parameters are the displacement phases $\varphi_i$ which can now be determined by $g^{(2)}$ measurements. Having a multimode displaced squeezed thermal state, the second-order correlations read,
\begin{align}
 g^{(2)}_{ij} &= 1 + |g_{ij}^{(1)}|^2 + \frac{|\cov{ij}|^2 - |\alpha_i|^2|\alpha_j|^2}{\bar{n}_i \bar{n}_j}  \notag \\ 
 &\quad + 2 \frac{|\alpha_i| \, |\alpha_j| \, |\cov{ij}|}{\bar{n}_i \bar{n}_j}  \cos(\Theta_{ij} - \varphi_i - \varphi_j).
\label{eq:g2MultiDSTexplicit}
\end{align}
From the diagonal elements $g^{(2)}_{ii}$, we can infer all $\varphi_i$ up to cosine ambiguities. These ambiguities can be resolved by including off-diagonal $g^{(2)}_{ij}$ measurements as these quantities explicitly include $\Theta_{ij}-\varphi_i-\varphi_j$ dependencies and implicitly relative phase dependencies $\varphi_i-\varphi_j$ via the $|g^{(1)}|^2$ contributions to Eq.~\eqref{eq:g2MultiDSTexplicit}, see Eq.~\eqref{eq:g1MultiDST}. These measurements also provide a nontrivial consistency check of whether a reconstruction in terms of a multimode displaced squeezed thermal state is possible. This can be further corroborated by additional measurements of $g^{(1)}$ and $g^{(3)}$. Alternatively, one may use measurements of $g^{(1)}_{ij}$ and Eq.~\eqref{eq:g1MultiDST} or analyzing $g^{(3)}_{iij}$ to infer all $\varphi_i$. Thus, we are able to provide evidence for and fully reconstruct a generic multimode displaced squeezed thermal state. Only in a two-mode scenario, we inherit the discrete four-fold ambiguity from the analysis of non-displaced Gaussian states in the zero-mean output.

Similarly, the alternative scheme that relies on the feasibility to manipulate the source of the Gaussian state which was presented for the single mode case (see Sec.~\ref{sec:SingleDST}), can in principle be extended to multimode systems. However, developing and experimentally implementing such a method is considerably more complex than in the single mode scenario. Thus, we do not pursue a detailed analysis here.

\section{Summary and Conclusion}\label{sec:Summary}
We have analyzed the connection between second- and third-order intensity correlation functions and photon-statistics-based state reconstruction for arbitrary (multi-mode) Gaussian states. The central technical ingredient is an explicit decomposition of sixth-order ladder-operator moments for Gaussian states, which allows us to express $g^{(3)}$ entirely in terms of first- and second-order moments and thereby derive general relations linking $g^{(3)}$ to $g^{(2)}$. These relations provide a practical tool for state characterization. They enable a hypothesis test of Gaussianity (violations certify non-Gaussianity) and, when satisfied, support a convenient classification of states into non-displaced, non-squeezed, and displaced-squeezed Gaussian sectors. Due to the normalization of the correlation functions $g^{(n)}$, this classification is robust against linear loss. While dark counts and other noise sources can in principle distort correlations and therefore spoil this classification, in typical photonic-based quantum information processing experiments these effects are usually negligible compared to optical loss which is the dominant imperfection~\cite{slussarenko_photonic_2019,you_superconducting_2020,ann_correction_2015}. Thus, the proposed scheme provides evidence for a Gaussian-state description and enables reliable classification into the aforementioned sectors. As discussed in Sec.~\ref{sec:SingleMode}, one could always construct pathological counterexamples of non-Gaussian states matching the correlation function dependence, however, this is ambiguity is inherit to any finite measurement set.

While the $g^{(2)}$-$g^{(3)}$ relations are loss-invariant, a crucial result of our analysis is that state reconstruction requires at least one loss-sensitive observable per mode, since Gaussian states remain Gaussian under linear loss. Concretely, supplementing intensity-correlation data with either the mean photon numbers $\bar{n}_i$ or (for click detection) the vacuum overlaps suffices to pin down the state parameters quantitatively. As an illustration, intensity correlations can certify that a single-mode state is non-displaced via the relation $g^{(3)}=9g^{(2)}-12$, yet they cannot discriminate a squeezed vacuum from a squeezed thermal state without $\bar n$ (or click probability) information.

For single-mode systems, we explicitly demonstrated that non-displaced Gaussian states or non-squeezed Gaussian states are fully reconstructible, up to a global phase-space rotation reflecting the gauge-like freedom to choose the coordinate system. For displaced squeezed Gaussian states, an unavoidable $\mathbb{Z}_2$ ambiguity remains mirroring the noise ellipse through the displacement axis, if we use only passive linear optics and intensity/click detection. If possible, it may be lifted by a parameter scan, e.g., varying the relative phase between displacement and squeezing (or by introducing a phase reference).

In the multimode setting, we showed that full state reconstruction is also possible for non-displaced or non-squeezed states up to a global phase rotation. In particular, the phases of displacement or squeezing parameters can be inferred from correlation data alone as they are unaffected by photon loss. We established the following recipe to extract the relevant state information for these two classes:
\begin{enumerate}
  \item Measure $\bar{n}_i$, $g_{ij}^{(1)}$, $g_{ij}^{(2)}$, and $g_{iij}^{(3)}$ for all contributing modes.
  \item Extract relative phase information via the functional dependence $g^{(3)}\!\big(g^{(2)},g^{(1)}\big)$.
  \item Use $\bar{n}_i$, $g^{(1)}_{ij}$, and $g^{(2)}_{ij}$ to determine all moduli $|\alpha_i|$ or $|\mathrm{cov}_{ij}|$.
  \item Reconstruct displacements/covariances up to a global phase.
  \item \emph{a) Displacement:} retrieve thermal excitations and mode rotations via the spectral decomposition of $\E^{i\boldsymbol\phi}\boldsymbol{\mathcal D}\E^{-i\boldsymbol\phi}$.\\
        \emph{b) Squeezing:} retrieve thermal excitations, mode rotations, and squeezing parameters via the Williamson decomposition of the real covariance matrix.
\end{enumerate}
Apart from the global phase rotation, an additional $\mathbb{Z}_2$ ambiguity may arise in this state reconstruction, though it is non-generic and can be neglected in typical experimental settings. The only exceptions are given by two-mode non-displaced states for which a residual four-fold discrete ambiguity persists and two-mode non-squeezed states with a residual two-fold discrete ambiguity. For a Gaussian state being displaced and squeezed, interference with an identically prepared state on a balanced beam splitter provides a straightforward route for state reconstruction. The zero-mean output yields squeezing and thermal properties via the above recipe, while the displacement parameters can be inferred from the other output.

Our discussion covered both intensity and click measurements to obtain the correlation functions and the mean photon number or simple click statistics. 
In principle, photon-number-resolving detection could be used equivalently to either access the required normally ordered moments or to derive the parameters characterizing the Gaussian state and performing evidence checks via measuring actual photon number probabilities. Practical experimental implementations of such protocols may be within reach due to recent advances in superconducting nanowire detector readout \cite{sauer_resolving_2023,schapeler_electrical_2024}. Further, the developed approach relies on access to mode-resolved information about the state. If detectors are mode-blind, i.e., measurements average over different modes, proper statements about state properties can only be obtained via additional assumptions. We discuss such a scenario in App.~\ref{app:BucketDetector}, considering a bucket detector without any modal resolution. An interesting and natural intermediate regime is coarse-grained detection, where detectors are blind within mode groups but distinguish between groups. In such cases, one may treat each group as an effective averaged mode and derive group-level relations among $g^{(1)}$, $g^{(2)}$, and $g^{(3)}$ from the presented sixth-order decomposition. 


In conclusion, the discussed photon-statistics-based methods for state classification and characterization provide an alternative to conventional Gaussian-state tomography that relies on local oscillators to define phase references. In particular, they may offer experimentally feasible routes to scalable and loss-robust characterization of multimode Gaussian resources.

\begin{acknowledgments}
We are grateful to Hendrik Ellenberg, David Theidel, and Ilya Karuseichyk for valuable discussions. We acknowledge funding by the Deutsche Forschungsgemeinschaft (DFG, German Research Foundation) -- Project-ID 398816777 -- SFB 1375 (project A7).
\end{acknowledgments}

\appendix

\section{Decomposition of Gaussian moments}\label{app.A}
The derivation of the necessary equations is done using the same techniques and methods as in \cite{vallone_means_2019}. By connecting the operator moments of a general Gaussian state to the moments of just thermal states through Bogoliubov transformations, a general sixth-order thermal moment can be used to calculate the respective sixth-order Gaussian moment.

In the main text, we always refer to an expectation value of a general Gaussian state when working with moments $\mathcal{O}_{\mu_1,\cdots,\mu_n}$ (Eqs.~\eqref{eq.11} and after). However, here we will denote whether we are dealing with just thermal states ($\mathrm{th}$) or general Gaussian states ($\mathrm{g}$) with a superscript for clarity.

In general, any thermal moment may be decomposed into a sum of products of second-order moments. To prove this, consider the following: It is easy to show that any odd moments vanish for an $M$-mode thermal state as there is always at least one ladder operator that appears without its conjugate. Hence, the number states in the thermal distribution do not match up anymore and the expectation value vanishes
\begin{align}\label{a1}
    \mot{\mu_1 \mu_2\cdots \mu_{2k+1}} = 0,
\end{align}
with $\mu_n \in \{\us i_1,\cdots, \us i_{2k+1}, i_1, \cdots, i_{2k+1}\}$ and $\mathcal{O}$ as in Eq.~\eqref{eq.11}. 
From this, we can immediately conclude that for even moments the operators need to come in pairs of creation and annihilation operators within each mode. Otherwise, the matching condition is again violated and the moment vanishes.

Due to the $M$-mode thermal state being a product state in terms of its modes (see the definition in Eq.~\eqref{eq:thermalState}), moments containing operators from differing modes separate,
\begin{equation}\label{a2}
\begin{aligned}
    \mot{i j \us i \us j} &= \mot{i \us i}\mot{j \us j} \quad \mathrm{for}\quad i\neq j.
\end{aligned}
\end{equation}
Of course, the order of the indices must still be respected. However, we can use a reordering scheme for $\mo{}$ (which holds true in general) using the commutation relation Eq.~\eqref{eq.1},
\begin{equation}\label{a3}
    \begin{aligned}
        \mot{ l k \us j \us i} &= \mot{ k l \us i \us j} =\mot{\us i k l \us j} + \delta_{il}\mot{k \us j} + \delta_{ik}\mot{l \us j} \\
        &= \mot{\us i \us j k l} + \delta_{il}\mot{\us j k} + \delta_{ik}\mot{\us j l} + \delta_{jl}\mot{\us i k}\\
        &\quad + \delta_{jk}\mot{\us i l} + \delta_{il}\delta_{jk} + \delta_{ik}\delta_{jl},
    \end{aligned}
\end{equation}
where normal (underlined) indices commute with each other whereas commuting underlined with normal indices results in an additional lower-order moment with a Kronecker delta in the commuted modes.

Now, any moment can be normal-ordered at the cost of introducing lower-order moments as well. Furthermore, it is well established that the normal-ordering of multiple number operators (in the same mode) leads to the photon number factorial moment
\begin{align}\label{a4}
        \langle :(\hat{a}^\dagger_i \hat{a}_i)^k:\rangle &= \langle:\hat{n}^k_i:\rangle \\
        &=\langle \hat{n}_i(\hat{n}_i-1)(\hat{n}_i-2)\cdots(\hat{n}_i-k+1)\rangle, \notag
\end{align}
which, for thermal states, yields \cite{loudon_the_1983}
\begin{equation}\label{a5}
        \langle:\hat{n}^k_i:\rangle = k!\langle\hat{n}_i\rangle^k 
       \implies \mot{\us i_1 \cdots \us i_k i_1 \cdots i_k} = k! (\mot{\us i i})^k,
\end{equation}
effectively reducing the order of the moment that needs to be computed. Since any moment can be normal-ordered by adding lower-order moments and number operators of different modes factor, it is apparent that any higher-order moment of a thermal state can be written as a sum of products of second-order moments. \\ \indent
In fact, the thermal moments for any combination of ladder operators can be evaluated by constructing a generalized version of \eqref{a5} and \eqref{a2}:
\begin{equation}\label{a6}
    \begin{aligned}
        \mot{\mu_1\mu_2\cdots\mu_{2n}} &= \sum_{s\in\mathcal{P}^n_2(\omega)}\prod_{i=1}^n \mot{s_{i1} s_{i2}}.
    \end{aligned}
\end{equation}
Here, $\mathcal{P}^n_2(\omega)$ refers to all partitions that divide the set of indices $\omega = \{\mu_1,\cdots,\mu_{2n}\}$ into $n$ subsets of size $2$. The indices $s_{ij} \in \omega$ are dependent on the given partition $s$ and specify the chosen indices for a given subset such that $\omega = \{s_{11},s_{12},\cdots, s_{n1},s_{n2}\}$ for all $s$. The number of addends in the sum grows with $n$ like 
\begin{equation}\label{a7}
    \#\mathcal{P}_2^n(\mu) = \frac{(2n)!}{2^n n!}.
\end{equation}
To better illustrate what is meant by this formula, let us look into two examples. The trivial case, where $n=1$, only has 1 possible partition as this is already just a second-order moment. The sum and product vanish and leave the moment unchanged. The first non-trivial case is the fourth-order moment $\mo{\mu\nu\kappa\lambda}$. The number of addends amounts to $\frac{4!}{2^2  2!} = 3$, meaning there are 3 unique partitions of the set $\omega = \{\mu,\nu,\kappa,\lambda\}$: $s^{(1)} = \qty{\{\mu,\nu\},\{\kappa,\lambda\}}$, $s^{(2)}=\qty{\{\mu,\kappa\},\{\nu,\lambda\}}$, and $s^{(3)} = \qty{\{\mu,\lambda\},\{\nu,\kappa\}}$. Lastly, the indices $s_{ij}$ are given by:
\begin{equation}\label{a8}
    \begin{aligned}
        s^{(1)} &=  \big\{\{\mu,\nu\},\{\kappa,\lambda\}\big\} = \big\{\{ s^{(1)}_{11}, s^{(1)}_{12}\},\{ s^{(1)}_{21}, s^{(1)}_{22}\}\big\}, \\
        s^{(2)} &=  \big\{\{\mu,\kappa\},\{\nu,\lambda\}\big\} =  \big\{\{ s^{(2)}_{11}, s^{(2)}_{12}\},\{ s^{(2)}_{21}, s^{(2)}_{22}\}\big\}, \\
        s^{(3)} &=  \big\{\{\mu,\lambda\},\{\nu,\kappa\}\big\} =  \big\{\{ s^{(3)}_{11}, s^{(3)}_{12}\},\{ s^{(3)}_{21}, s^{(3)}_{22}\}\big\}.
    \end{aligned}
\end{equation}
Since the moments are always split into a product of second-order moments, the second index of $s_{ij}$ can only be either 1 or 2. The first index can take values $i\in\{1,\cdots,n\}$. Consequently, the fourth-order moment is decomposed into:
\begin{equation}\label{a9}
    \begin{aligned}
        \mot{\mu\nu\kappa\lambda} &= \mot{\mu\nu}\mot{\kappa\lambda} + \mot{\mu\kappa}\mot{\nu\lambda} + \mot{\mu\lambda}\mot{\nu\kappa},
    \end{aligned}
\end{equation}
which corresponds to the result found in \cite{vallone_means_2019} for thermal states.

In particular, this decomposition is applicable to the sixth-order moments required for calculating the third-order coherence functions. In order to find an analogous formula for the sixth-order moment and to further motivate Eq.~\eqref{a6}, consider the moment in the third-order correlation function but with a thermal expectation value. This moment can be calculated straightforwardly and yields
\begin{equation}\label{a10}
    \begin{aligned}
        \mot{\us i \us j\us k i j k} &=  N_i N_j N_k + \delta_{ij}N_k N_j^2 + \delta_{jk}N_i N_k^2\\
        &\quad + \delta_{ik}N_jN_i^2 + 2\delta_{ij}\delta_{jk}N_i^3,
    \end{aligned}
\end{equation}
with $N_i$ being the expected thermal photon number in mode $i$. From this moment, any other non-vanishing, sixth-order moment can be calculated by changing the order of indices using the established commutation rules \eqref{a3} and results in a similar expression but with potential changes as $N_i \to N_i+1$.

With this in mind, Eq.~\eqref{a10} can be rewritten in terms of second-order moments
\begin{equation}\label{a11}
    \begin{aligned}
         \mot{\us i \us j\us k i j k} &= 
         \mot{\us i i}\mot{\us j j}\mot{\us k k} + \mot{\us i j}\mot{\us j i}\mot{\us k k} + \mot{\us j k}\mot{\us k j}\mot{\us i i} \\
         & \quad + \mot{\us i k}\mot{\us k i}\mot{\us j j} + \mot{\us i j}\mot{\us j k}\mot{\us k i} + \mot{\us i k}\mot{\us j i}\mot{\us k j},
    \end{aligned}
\end{equation}
with $\mot{\us i j} = \delta_{ij}N_i$. The effect of swapping the position of indices is now rather obvious as $\mot{j \us i} = \delta_{ij}(N_i + 1)$.

However, this does not capture the entire picture as we already made sure to match underlined and normal indices. In order to account for arbitrary indices, we can make sure to include all possible partitions of the sixth-order moment into a set of 3 second-order moments. In case of Eq.~\eqref{a11}, this would result in 9 additional terms that all vanish but would read:
\begin{equation}\label{a12}
    \begin{aligned}
         \mot{\us i \us j\us k i j k} &= 
         \phantom{+}\, \mot{\us i i}\mot{\us j j}\mot{\us k k} + \mot{\us i j}\mot{\us j i}\mot{\us k k} + \mot{\us j k}\mot{\us k j}\mot{\us i i} \\
         & \quad + \mot{\us i k}\mot{\us k i}\mot{\us j j} + \mot{\us i j}\mot{\us j k}\mot{\us k i} + \mot{\us i k}\mot{\us j i}\mot{\us k j}\\
         & \quad + \mot{\us i \us j}\mot{\us k i}\mot{j k} + \mot{\us i \us j}\mot{\us k j}\mot{i k} + \mot{\us i \us j}\mot{\us k k}\mot{i j}\\
         & \quad + \mot{\us i \us k}\mot{\us j i}\mot{j k} + \mot{\us i \us k}\mot{\us j j}\mot{i k} + \mot{\us i \us k}\mot{\us j k}\mot{i j}\\
         & \quad + \mot{\us j \us k}\mot{\us i i}\mot{j k} + \mot{\us j \us k}\mot{\us i j}\mot{i k} + \mot{\us j \us k}\mot{\us i k}\mot{i j}.
    \end{aligned}
\end{equation}
Here, any possible combinations are being accounted for and the order can easily be swapped on both sides of the equation which allows to loosen the matching condition and replace the six indices with arbitrary ones:
\begin{equation}\label{a13}
    \begin{aligned}
         \mot{\mu\nu\kappa\lambda\rho\sigma} &=  \phantom{+}\, 
         \mot{\mu\nu}\mot{\kappa\lambda}\mot{\rho\sigma} + \mot{\mu\nu}\mot{\kappa\rho}\mot{\lambda\sigma} + \mot{\mu\nu}\mot{\kappa\sigma}\mot{\lambda\rho} \\
         & \quad + \mot{\mu\kappa}\mot{\nu\lambda}\mot{\rho\sigma} + \mot{\mu\kappa}\mot{\nu\rho}\mot{\lambda\sigma} + \mot{\mu\kappa}\mot{\nu\sigma}\mot{\lambda\rho}\\
         & \quad + \mot{\mu\lambda}\mot{\nu\kappa}\mot{\rho\sigma} + \mot{\mu\lambda}\mot{\nu\rho}\mot{\kappa\sigma} + \mot{\mu\lambda}\mot{\nu\sigma}\mot{\kappa\rho}\\
         & \quad + \mot{\mu\rho}\mot{\nu\kappa}\mot{\lambda\sigma} + \mot{\mu\rho}\mot{\nu\lambda}\mot{\kappa\sigma} + \mot{\mu\rho}\mot{\nu\sigma}\mot{\kappa\lambda}\\
         & \quad + \mot{\mu\sigma}\mot{\nu\kappa}\mot{\lambda\rho} + \mot{\mu\sigma}\mot{\nu\lambda}\mot{\kappa\rho} + \mot{\mu\sigma}\mot{\nu\rho}\mot{\kappa\lambda}.
    \end{aligned}
\end{equation}
Equation \eqref{a13} has now been constructed from a single sixth-order moment and does, in fact, match up with the decomposition formula in Eq.~\eqref{a6}. The number of addends amounts to $\#\mathcal{P}^3_2(\mu) = \frac{6!}{8 \cdot 3!} = 15$ many of which vanish when computing a particular moment as for Eq.~\eqref{a11}.

Using this decomposition formula, the thermal expectation value can be transformed into a general Gaussian expectation value by applying the corresponding general Bogoliubov transformation Eq.~\eqref{eq.6}:
\begin{equation}\label{a14}
    \begin{aligned}
        \mog{\mu}&=\langle \hat{b}_\mu\rangle_\mathrm{g} = \mathrm{Tr}\big[\hat{b}_\mu\hat{\rho}_g\big]
        = \mathrm{Tr}\big[\hat{b}_\mu\hat{U}\hat{\rho}_\mathrm{th}\hat{U}^\dagger\big]\\
        &= \mathrm{Tr}\big[\hat{U}^\dagger\hat{b}_\mu\hat{U}\hat{\rho}_\mathrm{th}\big]
        = \mathrm{Tr}\big[(L_{\mu}^{\phantom{\mu}\nu}\hat{b}_\nu + A_\mu)\hat{\rho}_\mathrm{th}\big]\\
        &= \langle L_{\mu}^{\phantom{\mu}\nu} \hat{b}_\nu + A_\mu\rangle_\mathrm{th} 
        = L_\mu^{\phantom{\mu}\nu}\mot{\nu} + A_\mu,
    \end{aligned}
\end{equation}
where we used the cyclicity of the trace operation. From this, it can be concluded that $\mog{\mu} = A_\mu$ as the first-order thermal moments vanish. Analogously, second-, third- and fourth-order moments of general Gaussian states have been considered in \cite{vallone_means_2019} and read:
\begin{subequations}\label{a15}
    \begin{align}
        \mog{\mu}                &= A_\mu\label{a15a}\\
        \mog{\mu\nu}             &= L_\mu^{\phantom{\mu}\mu'}L_\nu^{\phantom{\nu}\nu'}\mot{\mu'\nu'} + A_\mu A_\nu\label{a15b}\\
        \mog{\mu\nu\kappa}       &= \mog{\mu\nu}A_\kappa + \mog{\mu\kappa}A_\nu + \mog{\nu\kappa}A_\mu  \notag \\
                                 &\quad - 2 A_\mu A_\nu A_\kappa  \label{a15c} \\
        \mog{\mu\nu\kappa\lambda}&= \mog{\mu\nu}\mog{\kappa\lambda}+\mog{\mu\kappa}\mog{\nu\lambda}+\mog{\mu\lambda}\mog{\nu\kappa} \notag \\
                                 &\quad  - 2A_\mu A_\nu A_\kappa A_\lambda, \label{a17ad}
    \end{align}
\end{subequations}
where the first and second equation were used to express the third- and fourth-order moments in terms of Gaussian moments as well.

Similarly, we can calculate the Gaussian sixth-order moments:
\begin{align}
    \mog{\mu\nu\kappa\lambda\rho\sigma} &=
    \sm{\mu}\sm{\nu}\sm{\kappa}\sm{\lambda}\sm{\rho}\sm{\sigma}\mot{\mu'\nu'\kappa'\lambda'\rho'\sigma'} \notag \\
    &\quad + \!\!\!\! \sum_{\substack{\{\psi,\chi\} \\ \in\mathcal{P}_{4,2}(\omega)}}        
        \sm{\psi_1}\sm{\psi_2}\sm{\psi_3}\sm{\psi_4}\mot{\psi_1'\psi_2'\psi_3'\psi_4'} A_{\chi_1}A_{\chi_2} \notag \\
    &\quad + \!\!\!\! \sum_{\substack{\{\psi,\chi\} \\ \in\mathcal{P}_{4,2}(\omega)}}        
        \sm{\chi_1}\sm{\chi_2}\mot{\chi_1'\chi_2'} A_{\psi_1}A_{\psi_2}A_{\psi_3}A_{\psi_4} \notag \\
    &\quad +  A_\mu A_\nu A_\kappa A_\lambda A_\rho A_\sigma,
\label{a16}
\end{align}
where $\mathcal{P}_{4,2}(\omega)$ refers to the bipartitions of the set of indices $\omega=\{\mu,\nu,\kappa,\lambda,\rho,\sigma\}$ into two disjoint subsets: one of size 4 ($\psi$) and another one of size 2 ($\chi$). For example, one such partition is $\{\psi,\chi\}$ with $\psi = \{\mu,\nu,\kappa,\lambda\}$ and $\chi = \{\rho,\sigma\}$. The indices in $\psi_i, \chi_i$ specify the elements within a given subset $\psi$ and $\chi$.

Necessarily, the ordering needs to be respected when constructing the partitions as the terms are a result of a product of creation and annihilation operators. There are $\binom{6}{2}=15$ such partitions in both sums in Eq.~\eqref{a16}.

As a next step, we can replace the fourth- and sixth-order thermal moments with their decompositions using Eq.~\eqref{a6} and rewrite the second-order moments as Gaussian moments with Eq.~\eqref{a15b}:
\begin{subequations}\label{a17}
    \begin{align}
        &\sm{\mu}\sm{\nu}\sm{\kappa}\sm{\lambda}\sm{\rho}\sm{\sigma}\mot{\mu'\nu'\kappa'\lambda'\rho'\sigma'} \notag \\
        &= \sum_{s\in\mathcal{P}^3_2(\omega)}\prod_{i=1}^3 \big[\mog{s_{i1} s_{i2}} - A_{s_{i1}}A_{s_{i2}}\big], \label{a17a} \\
        &\sm{\psi_1}\sm{\psi_2}\sm{\psi_3}\sm{\psi_4}\mot{\psi_1'\psi_2'\psi_3'\psi_4'} \notag \\
        &= \sum_{s\in\mathcal{P}^2_2(\psi)}\prod_{i=1}^2 \big[\mog{s_{i1} s_{i2}} - A_{s_{i1}}A_{s_{i2}}\big]. \label{a17b}
    \end{align}
\end{subequations}
Each of these terms can be inserted into Eq.~\eqref{a16}. We can now consider each term individually in order to find simplifications.

First, we rearrange Eq.~\eqref{a17a},
\begin{equation}\label{a18}
    \begin{aligned}
        &\sum_{s\in\mathcal{P}^3_2(\omega)}\prod_{i=1}^3 \big[\mog{s_{i1} s_{i2}} - A_{s_{i1}}A_{s_{i2}}\big] \\
        &= \!\!\sum_{s\in\mathcal{P}^3_2(\omega)}\prod_{i=1}^3 \mog{s_{i1} s_{i2}}
        - \!\!\!\!\sum_{s\in\mathcal{P}^3_2(\omega)}\sum_{i=1}^3 A_{s_{i1}} A_{s_{i2}} \prod_{j\neq i}^3 \mog{s_{j1} s_{j2}}\\
        &\quad  + \!\!\sum_{s\in\mathcal{P}^3_2(\omega)} \sum_{i=1}^3 \mog{s_{i1} s_{i2}} \prod_{j\neq i}^3 A_{s_{j1}} A_{s_{j2}}
         - \!\!\!\!\sum_{s\in\mathcal{P}^3_2(\omega)}\prod_{i=1}^3 A_{s_{i1}} A_{s_{i2}},
    \end{aligned}
\end{equation}
where the product is factored out such that we can order the terms depending on their number of second-order moments. The last term is fully symmetric in all partitions and consequently simplifies to
\begin{equation}\label{a19}
    \sum_{s\in\mathcal{P}^3_2(\omega)}\prod_{i=1}^3 A_{s_{i1}} A_{s_{i2}} = 15 A_\mu A_\nu A_\kappa A_\lambda A_\rho A_\sigma.
\end{equation}
Furthermore, the second and third term may lead to cancellations with other terms in Eq.~\eqref{a16}. In order to better identify these cancellations, we will adapt the way in which the summation is performed.

Consider, for example, the second term in Eq.~\eqref{a18}. In case of the 3 partitions $s^{(1)} = \{\{\mu,\nu\},\{\kappa,\lambda\},\{\rho,\sigma\}\}$, $s^{(2)} = \{\{\mu,\kappa\},\{\nu,\lambda\},\{\rho,\sigma\}\}$ and $s^{(3)} = \{\{\mu,\lambda\},\{\nu,\kappa\},\{\rho,\sigma\}\}$, the addends read
\begin{equation}\label{a20}
    \begin{aligned}
         &\big[A_\mu A_\nu \mog{\kappa \lambda}\mog{\rho\sigma} + A_\kappa A_\lambda \mog{\mu\nu} \mog{\rho\sigma} + A_\rho A_\sigma \mog{\mu\nu}\mog{\kappa\lambda}\big] \\
        +&\big[A_\mu A_\kappa \mog{\nu \lambda}\mog{\rho\sigma} + A_\nu A_\lambda \mog{\mu\kappa} \mog{\rho\sigma} + A_\rho A_\sigma \mog{\mu\kappa}\mog{\nu\lambda}\big]\\
        +&\big[A_\mu A_\lambda \mog{\nu \kappa}\mog{\rho\sigma} + A_\nu A_\kappa \mog{\mu\lambda} \mog{\rho\sigma} + A_\rho A_\sigma \mog{\mu\lambda}\mog{\nu\kappa}\big].
    \end{aligned}
\end{equation}
Here, we note that across those 3 partitions, the last terms within each line share a prefactor $A_\rho A_\sigma$, such that we can factor out this term,
\begin{equation}\label{a21}
    \begin{aligned}
        &A_\rho A_\sigma\big[\mog{\mu\nu}\mog{\kappa\lambda} + \mog{\mu\kappa}\mog{\nu\lambda}+\mog{\mu\lambda}\mog{\nu\kappa}\big] \\
        &= A_\rho A_\sigma     \sum_{s\in\mathcal{P}^2_2(\psi)}\prod_{i=1}^2\mog{s_{i1} s_{i2}},
    \end{aligned}
\end{equation}
and identify the remaining sum of products of moments as all possible partitions of the set $\psi = \{\mu,\nu,\kappa,\lambda\}$ into 2 subsets of size 2.

Analogously, any other combination of $A_{\omega_i} A_{\omega_j}$ can be factored from 3 corresponding moment products, such that all $\binom{6}{2} = 15$ combinations are represented. In turn, we can rewrite the double sum into a summation over all partitions $\{\psi,\chi\}\in\mathcal{P}_{4,2}(\omega)$ which split the set $\omega$ into a subset $\chi$ of size 2 (indices of the two $A$) and a subset $\psi$ of size 4 (indices of the two $\mog{}$), after which we sum over all partitions $s\in\mathcal{P}_2^2 (\psi)$ that split this $\psi$ into equal sized subsets:
\begin{align}
        &\sum_{s\in\mathcal{P}^3_2(\omega)}\sum_{i=1}^3 A_{s_{i1}} A_{s_{i2}} \prod_{j\neq i}^3 \mog{s_{j1} s_{j2}} \notag \\
        &\quad = \sum_{\substack{\{\psi,\chi\} \\ \in\mathcal{P}_{4,2}(\omega)}}  \sum_{s \in \mathcal{P}_2^2(\psi)} A_{\chi_1} A_{\chi_2} \prod_{i=1}^2 \mog{s_{i1} s_{i2}}.
\label{a22}
\end{align}
This may not seem helpful at first, but the utility of this restructuring of the sum will become evident below.

Similarly, we can re-sum the third term in Eq.~\eqref{a18}. Here, even more symmetries arise in the summation as the product $A_\mu A_\nu A_\kappa A_\lambda$ is fully symmetric under index permutation:
\begin{align}
        &\sum_{s\in\mathcal{P}^3_2(\omega)} \sum_{i=1}^3 \mog{s_{i1} s_{i2}} \prod_{j\neq i}^3 A_{_{j1}} A_{s_{j2}}  \notag \\
        &\quad = 3 \sum_{\substack{\{\psi,\chi\} \\ \in\mathcal{P}_{4,2}(\omega)}}     \mog{\chi_1 \chi_2} A_{\psi_1} A_{\psi_2}A_{\psi_3}A_{\psi_4}.
\label{a23}
\end{align}
As a last step, we need to consider Eq.~\eqref{a17b} as well. In analogy to Eq.~\eqref{a17a}, we separate the product into a sum:
\begin{align}
        &\sum_{s\in\mathcal{P}^2_2(\psi)}\prod_{i=1}^2 \big[\mog{s_{i1} s_{i2}} - A_{s_{i1}}A_{s_{i2}}\big] \notag\\
        &\quad = 
        \sum_{s\in\mathcal{P}^2_2(\psi)} \prod_{i=1}^2 \mog{s_{i1} s_{i2}}
        +  \sum_{s\in\mathcal{P}^2_2(\psi)} \prod_{i=1}^2 A_{s_{i1}} A_{s_{i2}} \notag\\
        &\qquad- \sum_{s\in\mathcal{P}^2_2(\psi)}\sum_{i=1}^2 \mog{s_{i1}s_{i2}} A_{s_{i1}} A_{s_{i2}}.
\label{a24}
\end{align}
The second term is fully symmetric and can therefore be calculated explicitly yielding
\begin{equation}\label{a25}
    \sum_{s\in\mathcal{P}^2_2(\psi)} \prod_{i=1}^2 A_{s_{i1}} A_{s_{i2}} = 3 A_{\psi_1} A_{\psi_2} A_{\psi_3} A_{\psi_4}.
\end{equation}
When applying the full summation from Eq.~\eqref{a16} to the second-to-last term, the double sum in the final term of Eq.~\eqref{a24} simplifies to
\begin{align}
        &\sum_{\substack{\{\psi,\chi\} \\ \in\mathcal{P}_{4,2}(\omega)}} \!\! A_{\chi_1}A_{\chi_2} \!\! \sum_{s \in \mathcal{P}_2^2(\psi)} \sum_{i=1}^2 \mog{s_{i1}s_{i2}} A_{s_{i1}} A_{s_{i2}} \notag \\
        &\qquad = 
        6 \sum_{\substack{\{\psi,\chi\} \\ \in\mathcal{P}_{4,2}(\omega)}}      \mog{\chi_1 \chi_2} A_{\psi_1}A_{\psi_2}A_{\psi_3}A_{\psi_4}.
\label{a26}
\end{align}
Now, we are able to fully evaluate Eq.~\eqref{a16}. Applying Eq.~\eqref{a15b} to the third addend and inserting the results from \cref{a17,a18,a19,a20,a21,a22,a23,a24,a25,a26} allows to fully express the sixth-order Gaussian moment only in term of first- and second-order Gaussian moments. The resulting decomposition yields
\begin{widetext}
\begin{align}
    \mog{\mu\nu\kappa\lambda\rho\sigma} &=
    \sum_{s\in\mathcal{P}^3_2(\omega)}\prod_{i=1}^3 \mog{s_{i1} s_{i2}}
    -     \sum_{\substack{\{\psi,\chi\} \\ \in\mathcal{P}_{4,2}(\omega)}}    \sum_{s \in \mathcal{P}_2^2(\psi)} A_{\chi_1} A_{\chi_2} \prod_{i=1}^2 \mog{s_{i1} s_{i2}}
    + 3  \!\!\sum_{\substack{\{\psi,\chi\} \\ \in\mathcal{P}_{4,2}(\omega)}}         \mog{\chi_1\chi_2}A_{\psi_1}A_{\psi_2}A_{\psi_3}A_{\psi_4} 
    \notag \\
    &\quad +  \sum_{\substack{\{\psi,\chi\} \\ \in\mathcal{P}_{4,2}(\omega)}}    \sum_{s \in \mathcal{P}_2^2(\psi)}         A_{\chi_1} A_{\chi_2} \prod_{i=1}^2 \mog{s_{i1} s_{i2}} - 6  \sum_{\substack{\{\psi,\chi\} \\ \in\mathcal{P}_{4,2}(\omega)}}       \mog{\chi_1 \chi_2} A_{\psi_1}A_{\psi_2}A_{\psi_3}A_{\psi_4} 
    \notag \\
    &\quad +   \sum_{\substack{\{\psi,\chi\} \\ \in\mathcal{P}_{4,2}(\omega)}}       \mog{\chi_1 \chi_2} A_{\psi_1}A_{\psi_2}A_{\psi_3}A_{\psi_4}
    + 16 A_\mu A_\nu A_\kappa A_\lambda A_\rho A_\sigma,
\label{a27}
\end{align}
\end{widetext}
where due to the re-summing it becomes evident that the terms quadratic in second-order moments cancel out completely. 
Additionally, terms linear in second-order moments can be combined as well leading to the final decomposition formula of the sixth-order Gaussian moment,
\begin{align}
        \mog{\mu\nu\kappa\lambda\rho\sigma} &= \sum_{s\in\mathcal{P}^3_2(\omega)}\prod_{i=1}^3 \mog{s_{i1} s_{i2}} + 16 A_\mu A_\nu A_\kappa A_\lambda A_\rho A_\sigma \notag \\
        &\quad -2         \sum_{\substack{\{\psi,\chi\} \\ \in\mathcal{P}_{4,2}(\omega)}}       \mog{\chi_1 \chi_2} A_{\psi_1}A_{\psi_2}A_{\psi_3}A_{\psi_4}. 
\label{a28}
\end{align}
This is the analogue to Eq.~\eqref{eq.13} for sixth-order moments. It consists of 15 terms cubic in second-order moments, 15 terms linear in second-order moments and quartic in first-order moments and a final term containing 6 first-order moments. Rewriting $A_\mu = \mog{\mu}$ to avoid excessive notational complexity leads to Eq.~\eqref{eq.15} of the main text.

%
Further, we can now start to work with correlation functions. The third-order correlation function in particular demands the sixth-order moment $\mog{\us i \us j \us k i j k}$, where $\omega = \{\us i, \us j, \us k, i, j, k\}$. 
Using Eq.~\eqref{a28} we can decompose this moment. When considering a general Gaussian state of the form in Eq.~\eqref{eq:multiGaussian}, we can simplify first-order moments: $\mog{m} = \alpha_m$, $\mog{\us m} = \alpha_m^*$ with $\alpha_m$ being the displacement in mode $m$. Hence, the sixth-order moment reads
\begin{equation}\label{b1}
    \begin{aligned}
        &\mog{\us i \us j \us k i j k}= 16 |\alpha_i|^2 |\alpha_j|^2 |\alpha_k|^2\\
        &\qquad\quad\begin{matrix}
        + \mog{\us i\us j}\mog{\us k i}\mog{jk} &+ \mog{\us i\us j}\mog{\us k j}\mog{ik} &+ \mog{\us i\us j}\mog{\us k k}\mog{ij}\\
        + \mog{\us i\us k}\mog{\us j i}\mog{jk} &+ \mog{\us i\us k}\mog{\us j j}\mog{ik} &+ \mog{\us i\us k}\mog{\us j k}\mog{ij}\\
        +\mog{\us i i}\mog{\us j\us k}\mog{jk} &+ \mog{\us i i}\mog{\us j j}\mog{\us k k} &+ \mog{\us i i}\mog{\us j k}\mog{\us k j}\\
        + \mog{\us i j}\mog{\us j\us k}\mog{ik} &+ \mog{\us i j}\mog{\us j i}\mog{\us k k} &+ \mog{\us i j}\mog{\us j k}\mog{\us k i}\\
        + \mog{\us i k}\mog{\us j\us k}\mog{ij} &+ \mog{\us i k}\mog{\us j j}\mog{\us k i} &+ \mog{\us i k}\mog{\us j i}\mog{\us k j}\\
        \end{matrix}\\
        &-2\left[\begin{matrix}
        \mog{\us i\us j} |\alpha_k|^2\alpha_i \alpha_j &+ \mog{\us i\us k} |\alpha_j|^2\alpha_i \alpha_k &+ \mog{\us j\us k} |\alpha_i|^2\alpha_j \alpha_k\\
        + \mog{ij} |\alpha_k|^2\alpha_i^* \alpha_j^* &+ \mog{ik} |\alpha_j|^2\alpha_i^* \alpha_k^* &+ \mog{jk} |\alpha_i|^2\alpha_j^* \alpha_k^*\\
        + \mog{\us i i} |\alpha_j|^2|\alpha_k|^2 &+ \mog{\us i j} |\alpha_k|^2\alpha_i \alpha_j^* &+ \mog{\us i k} |\alpha_j|^2\alpha_i \alpha_k^*\\
        + \mog{\us j i} |\alpha_k|^2\alpha_i^* \alpha_j &+ \mog{\us j j} |\alpha_i|^2|\alpha_k|^2 &+ \mog{\us j k} |\alpha_i|^2\alpha_j \alpha_k^*\\ 
        + \mog{\us k i} |\alpha_j|^2\alpha_i^* \alpha_k &+ \mog{\us k j} |\alpha_i|^2\alpha_j^* \alpha_k &+ \mog{\us k k} |\alpha_i|^2|\alpha_j|^2
        \end{matrix}\right]\\
    \end{aligned}
\end{equation}
The resulting terms can be combined to obtain real parts of complex numbers. Additionally, we normalize by the mean photon numbers and replace the second-order moments with first-order correlation functions and ladder operator covariances $\cov{ij}$ to find:
\begin{widetext}
\begin{align}
g^{(3)}_{ijk} &= 1 + |g^{(1)}_{ij}|^2 + | g^{(1)}_{jk} |^2 + | g^{(1)}_{ik} |^2 
+ 2 \, \mathrm{Re}[ g^{(1)}_{ij} g^{(1)}_{jk} g^{(1)}_{ki} ] 
+ \frac{  | \mathrm{cov}_{ij}   |^2 - | \alpha_i \alpha_j  |^2}{n_i n_j}
+ \frac{  | \mathrm{cov}_{jk}   |^2 - | \alpha_j \alpha_k   |^2}{n_j n_k}
+ \frac{  | \mathrm{cov}_{ik}   |^2 - | \alpha_i \alpha_k   |^2}{n_i n_k} 
\notag \\
&\quad +   \left( 2 - 4 \frac{|\alpha_k|^2}{n_k}   \right) 
  \frac{ \mathrm{Re}    [ \mathrm{cov}_{ij} \alpha_i^* \alpha_j^*   ]}{n_i n_j}
+   \left( 2 - 4 \frac{|\alpha_j|^2}{n_j}   \right) 
  \frac{ \mathrm{Re}    [ \mathrm{cov}_{ik} \alpha_i^* \alpha_k^*   ]}{n_i n_k} 
+   \left( 2 - 4 \frac{|\alpha_i|^2}{n_i}   \right) 
  \frac{ \mathrm{Re}    [ \mathrm{cov}_{jk} \alpha_j^* \alpha_k^*   ]}{n_j n_k} + 4 \frac{|\alpha_i \alpha_j \alpha_k|^2}{n_i n_j n_k}   
 \notag \\
&\quad + 2\frac{\mathrm{Re}    [ g^{(1)}_{ij} \, \mathrm{cov}^*_{jk} \mathrm{cov}_{ik}   ]}{n_k \sqrt{n_i n_j}}
+ 2\frac{\mathrm{Re}    [ g^{(1)}_{jk} \, \mathrm{cov}^*_{ik} \mathrm{cov}_{ij}   ]}{n_i \sqrt{n_j n_k}}
+ 2\frac{\mathrm{Re}    [ g^{(1)}_{ik} \, \mathrm{cov}^*_{ij} \mathrm{cov}_{jk}   ]}{n_j \sqrt{n_i n_k}} - 2 \frac{|\alpha_k|^2}{n_k} \frac{ \mathrm{Re}    [ g^{(1)}_{ij} \alpha_i \alpha_j^*   ]}{\sqrt{n_i n_j}}
\notag\\
&\quad - 2 \frac{|\alpha_i|^2}{n_i} \frac{ \mathrm{Re}    [ g^{(1)}_{jk} \alpha_j \alpha_k^*   ]}{\sqrt{n_j n_k}} 
- 2 \frac{|\alpha_j|^2}{n_j} \frac{ \mathrm{Re}    [ g^{(1)}_{ki} \alpha_k \alpha_i^*   ]}{\sqrt{n_i n_k}}
+ 2\frac{ \mathrm{Re}[ g^{(1)}_{ij} \alpha_i \alpha_k \mathrm{cov}^*_{jk}   ]+ \mathrm{Re}[ g^{(1)}_{ij} \alpha_j^* \alpha_k^* \mathrm{cov}_{ik}]}{n_k \sqrt{n_i n_j}} \notag \\
&\quad + 2\frac{\mathrm{Re}[ g^{(1)}_{jk} \alpha_i \alpha_j \mathrm{cov}^*_{ik}   ] + \mathrm{Re}[ g^{(1)}_{jk} \alpha_i^* \alpha_k^* \mathrm{cov}_{ij}]}{n_i \sqrt{n_j n_k}}
+ 2\frac{ \mathrm{Re}[ g^{(1)}_{ik} \alpha_j \alpha_k \mathrm{cov}^*_{ij}] + \mathrm{Re}[ g^{(1)}_{ik} \alpha_i^* \alpha_j^* \mathrm{cov}_{jk}]}{n_j \sqrt{n_i n_k}}.
\end{align}
\end{widetext}
Setting either $\cov{ij} = 0$ or $\alpha_i=0$ results in the versions for non-displaced or non-squeezed states in Eq.~\eqref{eq:g3multiDT} and Eq.~\eqref{eq:g3-g2multiST}, respectively.

\section{Phase reconstructions}\label{app:phases}
Here, we present a method of reconstructing the displacement (covariance) phases $\varphi_i$ ($\Theta_{ij}$) systematically using a graph-theoretical approach. Depending on the type of Gaussian state, the approach differs slightly, since more independent cases must be considered when dealing with squeezing due to the underlying matrix structure. However, the fundamental concept remains the same.

Let us start by considering the set of equations from which we aim to reconstruct the phases in the case of displacement. Rearranging Eqs.~\eqref{eq:giijDT} yields a set of equations of the form:
\begin{equation}\label{eq:Dsystem}
c_{ij} = \cos(\Phi_{ij} + \varphi_i - \varphi_j) \quad \text{with}\quad i\neq j,
\end{equation}
in which $c_{ij}$ and $\Phi_{ij}$ are known from correlation function measurements. Immediately, a global phase-shift symmetry is observed, as $\varphi_i\to\varphi_i + \psi_\alpha$ leaves all equations invariant. This global symmetry allows to set $\varphi_1 = 0$ without loss of generality. Doing so naturally splits the system into two types of equations:
\begin{subequations}
\begin{align}
c_{1i} &= \cos(\Phi_{1i} - \varphi_i)
\implies \varphi_i = \Phi_{1i} + \sigma_i\Tilde{c}_{1i},\label{eq:Dpi}\\
c_{ij} &= \cos(\Phi_{ij} + \varphi_i - \varphi_j) \quad i\neq j\label{eq:Dpipj},
\end{align}
\end{subequations}
where $\Tilde{c}$ abbreviates the $\arccos$ function acting on $c$ and $ \sigma_i\in \{-1,+1\}$ are binary variables representing the $\mathds{Z}_2$ symmetry of the cosine. Observe that for $\Tilde{c}_{1i} = 0$ a degeneracy occurs that removes the binary choice for $\varphi_i$. Importantly, such a degeneracy can only decrease the number of possible solutions, as we strictly decrease the number of degrees of freedom. We therefore confine ourselves, for the moment, to the case $\Tilde{c}_{1i}\neq 0$ (which holds for generic cases). In addition, by inserting Eq.~\eqref{eq:Dpi} into Eq.~\eqref{eq:Dpipj} and rearranging terms, we define:
\begin{subequations}
    \begin{align}
        \Gamma_{ij} &= \Phi_{ij} + \Phi_{1i} - \Phi_{1j} \\
        \Delta_{ij}(\sigma) &= \sigma_i \Tilde{c}_{1i} - \sigma_j \Tilde{c}_{1j}\\
        \xRightarrow{\eqref{eq:Dpipj}} c_{ij} &= \cos(\Gamma_{ij} + \Delta_{ij}(\sigma))\label{eq:Dvertex}.
    \end{align}
\end{subequations}
Consequently, we are dealing with $M-1$ individual phases in Eq.~\eqref{eq:Dpi}, each of which has a binary degree of freedom. However, the coupling equations in Eqs.~\eqref{eq:Dvertex} strongly restrict the number of possible solutions, as the signature $\sigma = \{\sigma_i\} $ must be consistent with all couplings. Identifying Eqs.~\eqref{eq:Dpi} or equivalently Eqs.~\eqref{eq:Dvertex} as edges connecting the vertices $\varphi_i$ (and $\varphi_1$), we obtain a (complete) graph of $M$ vertices. 
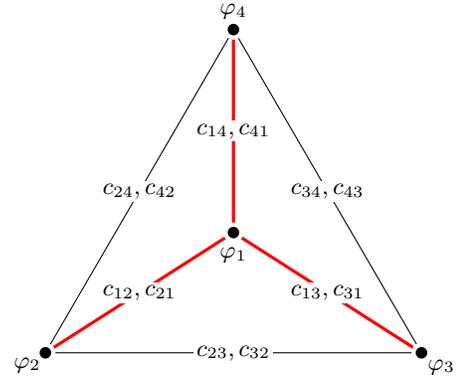
\begin{figure}[t]
    \centering
    \begin{tikzpicture}[edgelabel/.style={fill=white, inner sep=1pt}]
\coordinate (v1) at (0,0);
\coordinate (v2) at (5,0);
\coordinate (v3) at (2.5,4.3);
\coordinate (v4) at (2.5,1.6);
\draw (v1) -- node[midway, edgelabel] {$c_{23},c_{32}$} (v2);
\draw (v2) -- node[midway, edgelabel] {$c_{34},c_{43}$} (v3);
\draw (v1) -- node[midway, edgelabel] {$c_{24},c_{42}$} (v3);
\draw[red, very thick] (v4) -- node[midway, edgelabel, text=black] {$c_{12},c_{21}$} (v1);
\draw[red, very thick] (v2) -- node[midway, edgelabel, text=black] {$c_{13},c_{31}$} (v4);
\draw[red, very thick] (v3) -- node[midway, edgelabel, text=black] {$c_{14},c_{41}$} (v4);
\tikzset{dot/.style={circle, fill=black, draw=white, line width=1.6pt, minimum size=6pt, inner sep=0pt}}
\node[dot] (p1) at (v1) {};
\node[below left=-6pt of p1] {$\varphi_{2}$};
\node[dot] (p2) at (v2) {};
\node[below right=-5pt of p2] {$\varphi_{3}$};
\node[dot] (p3) at (v3) {};
\node[above=-3pt of p3] {$\varphi_{4}$};
\node[dot] (p4) at (v4) {};
\node[below=-1.5pt of p4] {$\varphi_1$};
\end{tikzpicture}
    \caption{Complete graph symbolizing the set of equations to determine the displacement phases for $M=4$. The global phase shift has been set to $\varphi_1=0$. Edges in red constitute a spanning tree of the graph.}
\label{fig:graph}
\end{figure}

Each edge corresponds to two equations which connect the corresponding vertices. In case of displacement, however, the equations for $c_{ij}$ and $c_{ji}$ have to yield the same result, meaning we can focus on $c_{ij}$ with $i<j$.

Now, solving the system proceeds in two steps: testing for the existence of solutions and characterize under which conditions multiple solutions may occur. The first step is performed by establishing a suitable spanning tree of the complete graph. In Fig.~\ref{fig:graph} a spanning tree connecting $\varphi_1$ to all other phases is chosen. Along this spanning tree, each choice of $\sigma_i$ is independent resulting in $2^{M-1}$ possible signatures $\sigma$. However, not all of them yield a unique or even valid solution set $\varphi = \{\varphi_i\}$ for the system. In order for $\sigma$ to generate a valid $\varphi$, all edge equations not part of the spanning tree need to be satisfied as well. Hence, we find a necessary and sufficient existence criterion: For a solution $\sigma$ to exist, it has to satisfy all Eqs.~$\eqref{eq:Dvertex}$. This is a highly restrictive condition which already renders the system for arbitrary sets $\Phi_{ij}$ and $c_{ij}$ without any solution. The existence of at least one admissible $\sigma$ therefore already constitutes a non-trivial test of Gaussianity.

On the other hand, if such a $\sigma$ exists, we can proceed to construct the phases $\varphi_i$ uniquely. There may, however, be multiple valid signatures. As we are dealing with binary choices in $\sigma_i$ a natural candidate for a second solution is $-\sigma$. In order for both solutions $\pm\sigma$ to exist simultaneously, Eqs.~\eqref{eq:Dvertex} need to be satisfied for both $\pm \sigma$. Flipping all signs in $\sigma$ results in $\Delta_{ij}(-\sigma) = -\Delta_{ij}(\sigma)$ which in turn requires:
\begin{equation}
    \begin{aligned}
       \cos(\Gamma_{ij} + \Delta_{ij}(\sigma)) &= \cos(\Gamma_{ij} - \Delta_{ij}(\sigma)) \\
       \iff \sin(\Gamma_{ij})\sin(\Delta_{ij}(\sigma)) &= 0,
    \end{aligned}
\end{equation}
imposing very strong conditions on the initial parameters $\Phi_{ij}, c_{ij}$. Specifically, on each non-spanning tree edge
\begin{equation}
    \begin{aligned}
        \Gamma_{ij} &\equiv 0 \mod{\pi}\\
        \lor \quad \Delta_{ij}(\sigma)&\equiv 0 \mod{\pi},
        \label{eq:p2sol}
    \end{aligned}
\end{equation}
has to hold, where the first condition places stringent constraints on the admissible $\Phi_{ij}$, while the second demands the measurements $c_{1i}$ and $c_{1j}$ to have equal magnitude $|c_{1i}|=|c_{1j}|$. For each edge $(i,j)$ outside the spanning tree, at least one of these two conditions must be satisfied for $\sigma$ and $-\sigma$ to generate two distinct solutions $\varphi$ simultaneously.

Now, consider an alternative solution distinct from $\pm\sigma$. Let $\tau \neq \pm \sigma$ be a second solution of the system. Then:
\begin{equation}
    \exists (i,j) \in \{2,\cdots,M\}^{2}: i\neq j, \tau_i = \sigma_i \land \tau_j = -\sigma_j.
\end{equation}
For these indices, the following relations hold:
\begin{equation}
    \begin{aligned}
        \Delta_{ij}(\sigma) + \Delta_{ij}(\tau) &= 2\sigma_i\Tilde{c}_{1i} \\
        \Delta_{ij}(\sigma) - \Delta_{ij}(\tau) &= -2\sigma_j\Tilde{c}_{1j}.
        \label{eq:pdeltas}
    \end{aligned}
\end{equation}
Again, Eqs.~\eqref{eq:Dvertex} must hold for both solutions simultaneously. This permits relating the two solutions to each other and constrains the set of initial parameters for which both can exist:
\begin{subequations}\label{eq:pcases}
    \begin{align}
        \cos(\Gamma_{ij} + \Delta_{ij}(\sigma)) &= \cos(\Gamma_{ij} + \Delta_{ij}(\tau))\notag\\
        \implies \Delta_{ij}(\sigma) + \Delta_{ij}(\tau) &\equiv -2\Gamma_{ij} \,\mod{2\pi}\label{eq:pcase1} \\
        \lor\quad \Delta_{ij}(\sigma) -\Delta_{ij}(\tau)  &\equiv 0 \qquad \;\mod{2\pi}\label{eq:pcase2}.
    \end{align}
\end{subequations}
We now treat both cases separately. If Eq.~\eqref{eq:pcase2} holds, inserting Eq.~\eqref{eq:pdeltas} immediately yields $\Tilde{c}_{1j} \equiv 0 \mod{\pi}$. Conversely, under Eq.~\eqref{eq:pcase1}, substituting Eq.~\eqref{eq:pdeltas} gives $\sigma_i \Tilde{c}_{1i} \equiv -\Gamma_{ij} \mod{\pi}$, which\textemdash upon substitution into the standard vertex equation given by Eq.~\eqref{eq:Dvertex}\textemdash leads to
\begin{equation}
    \begin{aligned}
        c_{ij} &= \cos(\Gamma_{ij} + \sigma_i\Tilde{c}_{1i} - \sigma_j\Tilde{c}_{1j})\\
        &= \cos(-\sigma_{j}\Tilde{c}_{1j}) = c_{1j}.
    \end{aligned}
\end{equation}
Therefore, if $c_{1i} \neq \pm 1$, the edge values $c_{ij}$ and $c_{1j}$ have to coincide and  $c_{1i} = \cos(\Gamma_{ij})$. Consequently, there are 2 distinct scenarios for the index pair $(i,j)$ which we summarize as the set of necessary conditions $\mathcal{G}_{ij}$, of which at least 1 must hold to allow for the scenario
$\tau_i = \sigma_i \land \tau_j = -\sigma_j$:
\begin{equation}
    \mathcal{G}_{ij} = 
    \begin{cases}
        c_{1i} = \pm1  \qquad \lor & \\
        c_{1j} = c_{ij} \land c_{1i} = \cos(\Gamma_{ij})&
    \end{cases}
\end{equation}
However, the characterization of the system is not yet complete. We now consider a third index $k$, for which either $\tau_k = \sigma_k$ or $\tau_k = -\sigma_k$ holds. In the first case, we examine the edge $(j,k)$ and, by applying the same reasoning as before, obtain the additional constraint $\mathcal{G}_{jk}$. In the second case, we analyze the edge $(i,k)$ and derive the corresponding constraint $\mathcal{G}_{ik}$. In this way, for every pair of vertices where the binary choices $\sigma_{m/n}$ and $\tau_{m/n}$ differ locally, a condition $\mathcal{G}_{mn}$ arises.

Hence, whenever two distinct signatures $\tau \neq \pm\sigma$ simultaneously satisfy the system, a large set of constraints ${\mathcal{G}_{mn}}$ is induced. If this set is fulfilled, multiple signatures may be valid. However, these requirements are extremely restrictive, such that in nearly all generic situations the system admits at most one solution (often none). Additionally, only the second condition in $\mathcal{G}_{mn}$ truly results in distinct solutions $\varphi$ as $c_{1m} = \pm 1$ renders $\sigma_m$ obsolete and determines $\varphi_m$ uniquely.

The special case $\pm\sigma$ is less demanding, requiring only Eq.~\eqref{eq:p2sol} and thus permitting exactly two distinct solutions. By contrast, if all $\mathcal{G}_{mn}$ are satisfied, the number of valid solutions can become very large. In the maximally degenerate case (for example, when $c_{ij}=0$ for all $(i,j)$ and $\Gamma_{ij}=\pm\tfrac{\pi}{2}$) there are $2^{M-1}$ solutions, as every binary choice yields a distinct valid configuration. Since no $c_{1i}$ is degenerate in this case, each signature indeed produces a different solution vector $\varphi$.

For squeezed states, the procedure follows analogously with an important difference: here $c_{ij}$ and $c_{ji}$ may differ. Moreover, there is no direct relation between the phases $\Theta_{ii}$ and $\Theta_{jj}$. However, we can circumvent this issue by considering precisely $c_{ij}$ and $c_{ji}$ simultaneously:
\begin{equation}\label{eq:pS}
    \begin{aligned}
       c_{ij} &= \cos(\Phi_{ij} + \Theta_{ii}-\Theta_{ij}) \\
       &= \sqrt{1-\sin^2(\Phi_{ij} + \Theta_{ii}-\Theta_{ij})}\\
       \implies C_{ij}(\epsilon) &= \cos(2\Phi_{ij} + \Theta_{ii} - \Theta_{jj})
    \end{aligned}
\end{equation}
with $C_{ij}(\epsilon) = c_{ij} c_{ji} + \epsilon_{ij}\sqrt{(1-c_{ij}^2)(1-c_{ji}^2)}$ and $\epsilon_{ij}$ is an additional binary degree of freedom arising from algebraic manipulations.

This transforms the system to one closely resembling the displacement case with the only difference being the additional binary choices $\epsilon_{ij}$. After fixing $\Theta_{11} = 0$, the vertices and edges of the corresponding complete graph are given by
\begin{subequations}
    \begin{align}
        \Theta_{ii} &= 2\Phi_{1i} + \sigma_i \Tilde{C}_{1i}(\epsilon)\\
        C_{ij}(\epsilon) &= \cos(2\Gamma_{ij} + \Delta_{ij}(\sigma,\epsilon)),
    \end{align}
\end{subequations}
where $\Delta_{ij}(\sigma,\epsilon) = \sigma_i\Tilde{C}_{1i}(\epsilon) - \sigma_j\Tilde{C}_{1j}(\epsilon)$, as before but now with explicit dependence on the signature $\epsilon$. The subsequent analysis is analogous to the displacement case albeit somewhat more involved due to the additional binary degrees of freedom.

Once a solution $\{\Theta_{ii}\}$ has been constructed using the signatures $\sigma$ and $\epsilon$, the remaining angles $\Theta_{ij}$ are fully determined via Eq.~\eqref{eq:pS}:
\begin{equation}
    \begin{aligned}
        \sigma_{ij}\Tilde{c}_{ij} - \sigma_{ji}\Tilde{c}_{ji} &= 2\Phi_{ij} + \Theta_{ii} - \Theta_{jj},
    \end{aligned}
\end{equation}
since the signatures $\sigma_{ij}$ are fixed for a given set $\{\Theta_{ii}\}$. An ambiguity remains only in the degenerate case ${c_{ij}=0}$. However, if such a degeneracy occurs, multiple valid solutions $\{\Theta_{ii}\}$ exist by construction.

In conclusion, the phase reconstruction procedure serves not only as a non-trivial test for Gaussianity but also typically yields a unique set of phases for generic experimental data. Multiple solutions can only occur in special cases of highly symmetric and degenerate values of $c_{ij}$ and $\Phi_{ij}$. In such situations, no further unique information can be extracted from correlation function measurements.

\section{Mode-Insensitive (Bucket) Detection}
\label{app:BucketDetector}
In this appendix, we analyze the correlation measurements of Gaussian states with mode-insensitive bucket detectors, i.e., detectors that do not discriminate between any of the individual field modes. For intensity measurements, one can only extract $\langle \sum_i \hat{a}_i^\dagger \hat{a}_i \rangle = \sum_i \bar{n}_i$, i.e., the total photon number of the state, from such detectors. An ideal mode-insensitive click detector can be described by a binary positive operator valued measure given by the projector onto the multimode vacuum $\op{0}$ and the projector $\mathds{1}-\op{0}$ registering a click whenever at least one photon occupies any mode.

In such a scenario, we define the (normally ordered) $n$-th order bucket correlations as
\begin{align}
	g^{(n)}_{\mathrm{B}}  = \frac{\langle:\sum_{i_1} a_{i_1}^\dagger a_{i_1} \cdots \sum_{i_n} a_{i_n}^\dagger a_{i_n}: \rangle}{\langle\sum_i a_i^\dagger a_i \rangle^n}.
\end{align}
To express the second- and third-order bucket correlation function for general (multi-mode) Gaussian states in terms of first- and second-order moments, we use the fourth- and sixth-order ladder-operator moment decompositions derived in the main text. For the second-order bucket correlation function, we obtain
\begin{align}
	g^{(2)}_{\mathrm{B}} &= 1 + \frac{\Tr \mathbf{G}^2}{(\Tr \mathbf{G})^2} 
	+ \frac{\Tr(\boldsymbol{\cov{}}^\dagger\boldsymbol{\cov{}}) - |\vec{\alpha}|^4 }{(\Tr \mathbf{G})^2} \notag \\
	&\quad + 2 \frac{ \mathrm{Re} \big[\vec{\alpha}^\dagger \boldsymbol{\cov{}} \vec{\alpha}^* \big] }{(\Tr \mathbf{G})^2},
\end{align}
where $\mathbf{G}$ is a hermitian matrix with elements ${G^{(1)}_{ij} = \langle a_i^\dagger a_j\rangle}$ being the unnormalized first-order coherence functions. The diagonal elements of $\mathbf{G}$ correspond to the mean photon numbers in each mode, $G^{(1)}_{ii} = \bar{n}_i$, and $\Tr \mathbf{G} = \sum_i \bar{n}_i$ gives the total mean photon number. Further $\boldsymbol{\cov{}}$ is the covariance matrix with elements $\cov{ij}$ and $\vec{\alpha}$ is the displacement vector. 
Similarly, we obtain for the decomposition of the third-order bucket correlations
\begin{widetext}
\begin{align}
        g^{(3)}_{\mathrm{B}}   &=  1 + 3\frac{\Tr(\mathbf{G}^2)}{(\Tr\mathbf{G})^2} + 2\frac{\mathrm{Re}[\mathrm{Tr}(\mathbf{G}^3)]}{(\Tr\mathbf{G})^3} 
        + 3\frac{\Tr(\boldsymbol{\cov{}}^\dagger\boldsymbol{\cov{}}) - |\vec\alpha|^4}{(\Tr\mathbf{G})^2} 
        + 6\bigg(1-\frac{|\vec\alpha|^2}{\Tr\mathbf{G}}\bigg) \frac{\mathrm{Re}\big[\vec\alpha^\dagger\boldsymbol{\cov{}}\vec\alpha^*\big]}
        {(\Tr\mathbf{G})^2}
        + 4\frac{|\vec\alpha|^6}{(\Tr\mathbf{G})^3} \notag \\
        &\quad  + 6 \frac{\Tr(\boldsymbol{\cov{}}\, \mathbf{G}\, \boldsymbol{\cov{}}^\dagger)}{(\Tr\mathbf{G})^3} 
        - 6 \frac{|\vec\alpha|^2 \, \vec\alpha^\top\mathbf{G}\, \vec\alpha^* }{(\Tr\mathbf{G})^3} 
        + 12\frac{\mathrm{Re}[\vec\alpha^\top\mathbf{G}\, \boldsymbol{\cov{}}^\dagger\vec\alpha]}{(\Tr\mathbf{G})^3}.
\end{align}
\end{widetext}
Because a bucket detector averages across modes, phase relations and mode-specific structures that would appear in mode-resolved $g^{(2)}$ or $g^{(3)}$ measurements collapse into trace invariants (e.g., $\Tr\mathbf{G}^k$, $\Tr[\boldsymbol{\cov{}}^\dagger\boldsymbol{\cov{}}]$, $|\vec\alpha|^2$, $\Tr[\boldsymbol{\cov{}}\,\mathbf{G}\,\boldsymbol{\cov{}}^\dagger]$). Consequently, mode-insensitive measurements inevitably discard information, and state reconstruction is not possible from $g^{(n)}_{\mathrm{B}}$ data alone. Inferring even partial state information from $g^{(n)}_{\mathrm{B}}$ data requires restricting the state space via physically motivated assumptions. Such assumptions may be justified if the underlying mechanism of producing the state is known, e.g., generating squeezed states of light via spontaneous parametric down conversion (SPDC).

Suppose there is independent evidence (e.g., from the source model) that the state is Gaussian. Then it suffices to analyze the two classes of non-squeezed and non-displaced Gaussian states. Displaced squeezed Gaussian states can be reduced to the latter class by interfering the state with an identically prepared copy on a balanced beam splitter, which removes the displacement in one output port as discussed in the main text. 
In such a scenario, we are able to infer a nontrivial squeezing parameter if $g^{(2)}_{\mathrm{B}} > 2$. For non-squeezed states, $\Tr \mathbf{G}^2 - |\vec\alpha|^4 = \Tr \boldsymbol{\mathcal{D}}^2 + 2 \vec{\alpha}^\dagger \E^{\im\boldsymbol \phi}\boldsymbol{\mathcal{D}}\E^{-\im\boldsymbol \phi} \vec{\alpha} \geq 0$ as $\boldsymbol{\mathcal{D}}$ is a diagonal matrix with nonnegative elements such that $ \Tr \boldsymbol{\mathcal{D}}^2 \geq 0$ and $\vec{x}^\dagger \boldsymbol{\mathcal{D}} \vec{x} \geq 0$ for any $\vec{x}$ ($= \E^{-\im\boldsymbol \phi} \vec{\alpha}$). Further, using $\vec{x}^\dagger \boldsymbol{\mathcal{D}} \vec{x} \leq |\vec{x}|^2 \Tr \boldsymbol{\mathcal{D}}$, we have 
$\Tr \boldsymbol{\mathcal{D}}^2 + 2 \vec{\alpha}^\dagger \E^{\im\boldsymbol \phi}\boldsymbol{\mathcal{D}}\E^{-\im\boldsymbol \phi} \vec{\alpha}
\leq \Tr \boldsymbol{\mathcal{D}}^2 + 2 |\vec{\alpha}|^2 \Tr \boldsymbol{\mathcal{D}}
\leq (\Tr \boldsymbol{\mathcal{D}})^2 + 2 |\vec{\alpha}|^2 \Tr \boldsymbol{\mathcal{D}}
\leq (\Tr \boldsymbol{\mathcal{D}} + |\vec{\alpha}|^2)^2 = (\Tr \mathbf{G})^2$.
Thus,
\begin{align}
 1 \leq g^{(2)}_{\mathrm{B}} = 1 + \frac{\Tr \mathbf{G}^2 - |\alpha|^4}{(\Tr \mathbf{G})^2} \leq 2,
\end{align}
for any non-squeezed Gaussian state, as expected. Therefore, any $g^{(2)}_{\mathrm{B}} >2$ (or $g^{(2)}_{\mathrm{B}} <1$) certifies a nontrivial squeezing parameter (thus $\boldsymbol{\cov{}}\neq \mathbf 0$) if a Gaussian-state description is valid. However, this does not necessarily imply quadrature variances below shot noise, since a squeezed thermal state with sufficiently large thermal mean occupation can have all quadrature variances above the shot-noise limit. 
Furthermore, any $g^{(2)}_{\mathrm{B}} \leq 1$ implies that the Gaussian state must be squeezed and displaced as we have for non-displaced states ($\vec{\alpha}=\vec{0}$),
\begin{align}
 g^{(2)}_{\mathrm{B}} = 1 + \frac{\Tr \mathbf{G}^2 + \Tr(\boldsymbol{\cov{}}^\dagger\boldsymbol{\cov{}})}{(\Tr \mathbf{G})^2}\;\geq 1,
\end{align}
using $\Tr(\boldsymbol{\cov{}}^\dagger\boldsymbol{\cov{}})\geq 0$ and $\Tr \mathbf{G}^2 = \Tr (\mathbf{G}^\dagger \mathbf{G}) \geq 0$ where we exploited the hermiticity of $\mathbf{G}$.

Even if a nontrivial squeezing parameter can be inferred from a $g^{(2)}_{\mathrm{B}}$ measurement, the extraction of further state properties is obscured by the averaging procedure of the bucket detectors. By imposing additional, physically motivated constraints, e.g., assuming that the source produces pure states, our formalism reproduces established results from the literature, discussing frequency-domain measurement strategies for characterizing broadband squeezed vacuum \cite{christ_probing_2011}. 
In such a case, $\boldsymbol{\mathcal{D}} = 0$ and we may be able to perform a basis transformation of the modes to obtain uncorrelated twin- or single-beam squeezers (Schmidt decomposition of the signal and idler modes of SPDC). In case of twin-beam squeezers, it is possible to estimate the number of modes via measuring $g^{(2)}_{\mathrm{B}}$ \cite{christ_probing_2011}, if potential losses affect all modes equally. Unequal loss reweights the mode mixture, thus, changing the effective number of modes. In case the source produces multimode squeezed vacuum that can be transformed into uncorrelated single-mode squeezers, we have $\boldsymbol{z} = \mathrm{diag}(z_1,\cdots,z_M)$ and the correlation functions simplify substantially as $\Tr(\boldsymbol{\cov{}}^\dagger\boldsymbol{\cov{}}) = \Tr \mathbf{G}^2 +  \Tr \mathbf{G}$, $\Tr(\boldsymbol{\cov{}}^\dagger\boldsymbol{\cov{}} \mathbf{G}) = \Tr \mathbf{G}^3 +  \Tr \mathbf{G}^2$ with $\Tr \mathbf{G}^n = \sum_{i=1}^M \sinh^{2n}|z_i|$. Thus,
\begin{align}
 g^{(2)}_{\mathrm{B}} &= 1 + \frac{1}{\Tr \mathbf{G}} + 2\frac{\Tr \mathbf{G}^2}{(\Tr \mathbf{G})^2}, \\
 g^{(3)}_{\mathrm{B}} &= 1 + \frac{3}{\Tr \mathbf{G}} + \frac{6\Tr \mathbf{G}^2}{(\Tr \mathbf{G})^2} + \frac{6\Tr \mathbf{G}^2 + 8\Tr \mathbf{G}^3}{(\Tr \mathbf{G})^3}.
\end{align}
However, these results rely on strong modeling assumptions (Gaussian state, purity, equal loss across modes, ...) and must be interpreted within that framework. Any departures such as mode-dependent loss, residual correlations, or non-Gaussian noise can bias the inferred information. 
If narrowband filters are available and an effective mode discretization enables individual mode measurements, multimode approaches that avoid bucket detection across multiple contributing modes can provide significantly deeper insight into the underlying state structure.

\bibliographystyle{apsrev4-1}
\bibliography{g3g2paper_PRA}

\end{document}